\documentclass[a4paper,11pt]{article}
\usepackage{a4wide,amsmath,amssymb,bbm}
\usepackage[english]{babel}
\usepackage{hyperref,enumerate}
\usepackage{tikz} 
\hypersetup{
    colorlinks,
    linkcolor={black},
    citecolor={green!50!black},
    urlcolor={blue!50!black}
}

\newcommand{\Ad}{\mathrm{Ad}}

\makeatletter

\@addtoreset{equation}{section} \makeatother

\begin{document}

\hfill{Imperial-TP-LW-2016-03}

\vspace{30pt}

\begin{center}
{\LARGE{\bf Target space supergeometry of $\eta$ and $\lambda$-deformed strings}}

\vspace{50pt}

Riccardo Borsato \ \ and \ \ Linus Wulff

\vspace{15pt}

{\small {\it Blackett Laboratory, Imperial College, London SW7 2AZ, U.K.}
\\
\vspace{12pt}
\texttt{r.borsato, l.wulff@imperial.ac.uk}}\\

\vspace{100pt}

{\bf Abstract}
\end{center}
\noindent
We study the integrable $\eta$ and $\lambda$-deformations of supercoset string sigma models, the basic example being the deformation of the $AdS_5\times S^5$ superstring. 
We prove that the kappa symmetry variations for these models are of the standard Green-Schwarz form, and
we determine the target space supergeometry by computing the superspace torsion. 
We check that the $\lambda$-deformation gives rise to a standard (generically type II*) supergravity background; for the $\eta$-model 
the requirement that the target space is a supergravity solution 
translates into a simple condition on the $R$-matrix which enters the definition of the deformation. We further construct all such non-abelian $R$-matrices of rank four which solve the homogeneous classical Yang-Baxter equation for the algebra $\mathfrak{so}(2,4)$. 
We argue that most of the corresponding backgrounds are equivalent to sequences of non-commuting TsT-transformations, and verify this explicitly for some of the examples.

\pagebreak 
\tableofcontents

\setcounter{page}{1}


\section{Introduction and summary of results}
A remarkable property of the $AdS_5\times S^5$ superstring sigma model is its classical integrability~\cite{Bena:2003wd}, see~\cite{Arutyunov:2009ga} for a review. In fact, this property extends to several other symmetric space string backgrounds~\cite{Wulff:2014kja, Wulff:2015mwa}. Recently two interesting deformations of the $AdS_5\times S^5$ superstring sigma model\footnote{These deformations extend to any $\mathbbm{Z}_4$-symmetric supercoset sigma model, i.e. symmetric space RR string background preserving supersymmetry.} were proposed which preserve the integrability. The $\eta$-model~\cite{Delduc:2013qra} and $\lambda$-model~\cite{Hollowood:2014qma}, named after the corresponding deformation parameters. The former is based on the Yang-Baxter deformation of~\cite{Cherednik:1981df,Klimcik:2002zj,Klimcik:2008eq}, it generalises the case of  bosonic coset models~\cite{Delduc:2013fga}, and its essential ingredient is an $R$-matrix which satisfies the modified classical Yang-Baxter equation (MCYBE).  The $\lambda$-model was originally proposed by~\cite{Sfetsos:2013wia} and it extends the case of bosonic cosets~\cite{Hollowood:2014rla} (see also~\cite{Tseytlin:1993hm}). The construction is  based on a $G/G$ gauged  Wess-Zumino-Witten (WZW) model, and it is more naturally  interpreted as a deformation of the non-abelian T-dual of the original string.
The two deformations are closely related; in fact, in both cases the original symmetry algebra gets $q$-deformed~\cite{Delduc:2014kha,Hollowood:2015dpa} (with $q$ real and root of unity respectively), and the two models are related, at least at the classical level, by the Poisson-Lie T-duality of~\cite{Klimcik:1995ux,Klimcik:1995dy}, see~\cite{Vicedo:2015pna,Hoare:2015gda,Klimcik:2015gba}. 

The attempt of interpreting these deformations as string theories has raised interesting questions. 
In fact, both models possess a local fermionic symmetry believed to be the standard kappa symmetry---which was expected to guarantee a string theory interpretation. 
However, the target space of the  $\eta$-model derived in~\cite{Arutyunov:2013ega,Arutyunov:2015qva}\footnote{See~\cite{Hoare:2014pna} for lower dimensional examples of bosonic truncations and~\cite{Borsato:2016hud} for a review}  does not solve the type IIB supergravity equations~\cite{Arutyunov:2015qva}, but rather a generalisation of them as suggested in~\cite{Arutyunov:2015mqj}. These generalised equations ensure scale invariance for the sigma model, but are not enough to have the full Weyl invariance, which is present only when the target space satisfies the standard equations of supergravity.
For the $\lambda$-model, on the other hand, it was shown that the target space does solve the supergravity equations, at least in the case of $\lambda$-deformed $AdS_2\times S^2\times T^6$~\cite{Borsato:2016zcf} and $AdS_3\times S^3\times T^4$~\cite{Chervonyi:2016ajp} string sigma models\footnote{These results differ from the ones proposed in~\cite{Sfetsos:2014cea}. The metric in target space of the  $\lambda$-deformed $AdS_5\times S^5$ was obtained in~\cite{Demulder:2015lva}}.

A possible resolution for the puzzle posed by the $\eta$-model could have been that, after all, the possessed local fermionic symmetry was not the standard kappa symmetry of Green-Schwarz. 
However this state of affairs was clarified recently in~\cite{Wulff:2016tju} where it was shown that, contrary to what was commonly believed, kappa symmetry of the type II Green-Schwarz superstring does not imply the full equations of motion of type II supergravity.\footnote{Earlier indications of this was seen in the pure spinor string in \cite{Mikhailov:2012id}.} Rather it implies a weaker (generalized) version of these equations, whose bosonic subsector coincides with the equations written down in~\cite{Arutyunov:2015mqj}. These generalized supergravity equations involve a Killing vector field $K^a$, and reduce to the standard type II supergravity equations when this vector field is set to zero. 
This fact implies that kappa-symmetric backgrounds whose metric does not allow for isometries must in fact solve the standard type II equations. 
The $\lambda$-model falls into this class, which is consistent with the fact that the corresponding target spaces were found to be supergravity backgrounds.\footnote{We will actually see that the kappa symmetry transformations of the $\lambda$-model take the standard form only after inserting proper factors of $i$ (see sec. \ref{sec:kappa}). This leads to a target space geometry which is a solution of type II* rather than type II supergravity. In the case of $AdS_2\times S^2\times T^6$~\cite{Borsato:2016zcf} it was shown how one can get a standard (and real) type IIB background by analytic continuation, or equivalently by picking a different coordinate patch. The same should be true for the deformation of $AdS_3\times S^3\times T^4$~\cite{Chervonyi:2016ajp}, and probably $AdS_5\times S^5$.} 
On the other hand, the $\eta$-model typically leads to a target-space metric which possesses isometries, so that a priori it is not possible to exclude the possibility that it solves only the generalized supergravity equations. 
It should be mentioned that, given a solution of the generalized supergravity equations and provided that $K^a$ is space-like, it is possible to find a genuine supergravity solution which is formally T-dual to it~\cite{Hoare:2015wia,Arutyunov:2015mqj} (i.e. only at the classical level of the sigma model, ignoring the fact that the dilaton is linear in the coordinate along which T-duality is implemented). We will not consider this possibility here.

\vspace{12pt}

\textbf{\emph{Target space supergeometry}}

\noindent
The procedure for the $\eta$-deformation can be generalised\footnote{We will use the names ``$\eta$-deformation'' and ``Yang-Baxter deformation'' for both the homogeneous (CYBE) and inhomogeneous (MCYBE) cases, as we can treat them both at the same time.} also to the case when the $R$-matrix satisfies the classical Yang-Baxter equation (CYBE)~\cite{Kawaguchi:2014qwa,Matsumoto:2014cja,vanTongeren:2015soa}. Therefore several solutions exist and the question is which choices lead to a string theory, i.e. a target space that solves the standard type II supergravity equations. Here we will answer this question and find a simple (necessary and sufficient) condition on $R$. 
We will also determine the form of the target space (super) fields for both the $\eta$ and the $\lambda$-model in terms of the ingredients that define them (see section~\ref{sec:models} for their definition); we check that the models can be written in Green-Schwarz form and we work out the superspace torsion. The target space fields can then be read off by comparing to the expressions in~\cite{Wulff:2016tju,Wulff:2013kga}. This gives a simple way of extracting the target space backgrounds, much simpler than previous methods. The metric and B-field are easily read off directly from the sigma model Lagrangian, see~\eqref{eq:L-eta-lambda}. The NSNS three-form and RR fluxes are found to be given by the expressions\footnote{Note that here we write the  $\lambda$-model as a solution of type IIB supergravity, and the corresponding RR flux is imaginary. The background is real when written as a solution of type IIB*. The reason for this is a non-standard sign in the kappa symmetry transformations of the lambda model, see sec \ref{sec:kappa}.}
\begin{align}
H_{abc}=&\,3M_{[ab,c]}
-3i
\left\{\begin{array}{c}
\hat\eta^2\\
-\lambda^2\end{array}\right\}
M^{\hat\alpha2}{}_{[a}(\gamma_b)_{\hat\alpha\hat\beta}M^{\hat\beta2}{}_{c]}\,,
\\
\mathcal S^{\hat\alpha1\hat\beta2}
=&\,
8i\left\{\begin{array}{c}
[\Ad_h(1+2\hat\eta^{-2}-4\mathcal O_+^{-1})]^{\hat\alpha1}{}_{\hat\gamma1}\\
i\lambda[\Ad_h(1+\lambda(1-\lambda^{-4})\mathcal O_+^{-1})]^{\hat\alpha1}{}_{\hat\gamma1}
\label{eq:RR-intro}
\end{array}\right\}
\widehat{\mathcal K}^{\hat\gamma1\hat\beta2}\,,
\end{align}
where the upper (lower) expression in curly brackets refers to the $\eta$ ($\lambda$) model and $\hat\eta=\sqrt{1-c\eta^2}$. The RR field strengths are encoded in the bispinor defined as~\cite{Wulff:2016tju,Wulff:2013kga}
\begin{equation}
\mathcal S=-i\sigma^2\gamma^a\mathcal F_a-\frac{1}{3!}\sigma^1\gamma^{abc}\mathcal F_{abc}-\frac{1}{2\cdot5!}i\sigma^2\gamma^{abcde}\mathcal F_{abcde}\,,
\end{equation}
where for standard supergravity backgrounds $\mathcal F=e^{\phi}F$ contains the exponential of the dilaton. The remaining ingredients in these equations are defined in section \ref{sec:models}, in particular the operators $\mathcal O_+,\ M$ and the group element $h$ are defined in~\eqref{eq:Opm-lambda},~\eqref{eq:Opm-eta},~(\ref{eq:M}) and (\ref{eq:adh}). From our computation we obtain also  the Killing vector of the generalised type II equations
\begin{equation}
K^a=-\frac{i}{16}(\gamma^a)^{\hat\alpha\hat\beta}(\nabla_{\hat\alpha1}\chi_{\hat\beta1}-\nabla_{\hat\alpha2}\chi_{\hat\beta2})\,,
\end{equation}
where $\chi^I$ ($I=1,2$) are the would be dilatino superfields
\begin{equation}
\chi^1_{\hat\alpha}=
\frac{i}{2}
\left\{\begin{array}{c}
\hat\eta\\
-1\end{array}\right\}
\gamma^b_{\hat\alpha\hat\beta}[\Ad_hM]^{\hat\beta1}{}_b
\,,\qquad
\chi^2_{\hat\alpha}=
-\frac{i}{2}
\left\{\begin{array}{c}
\hat\eta\\
i\lambda\end{array}\right\}
\gamma^a_{\hat\alpha\hat\beta} M^{\hat\beta2}{}_a\,.
\end{equation}
When $K^a$ vanishes we have a standard supergravity solution and the dilaton is given by\footnote{For the $\lambda$-model this formula was argued in~\cite{Hollowood:2014qma}. It is also consistent with the form of the bosonic dilaton  suggested in~\cite{Kyono:2016jqy} for the $\eta$-model based on bosonic $R$-matrices.}
\begin{equation}
e^{-2\phi}=\mathrm{sdet}(\mathcal O_+)\,.
\end{equation}
For the $\lambda$-model $K^a$ automatically vanishes and the target space is always a supergravity solution, consistently with the observation of~\cite{Wulff:2016tju} and the previous findings~\cite{Borsato:2016zcf,Chervonyi:2016ajp}.

\vspace{12pt}

\textbf{\emph{The $\eta$-model as a string}}

\noindent
For the $\eta$-model the situation is more subtle. Let us review some details at this point and recall that the $\eta$-deformation is defined by an antisymmetric $R$-matrix on the algebra $R:\,\mathfrak g\rightarrow\mathfrak g$, $R^T=-R$, satisfying the (M)CYBE
\begin{equation}
[R(x),R(y)]-R([R(x),y]+[x,R(y)])=c[x,y]\,,\quad \forall x,y\in\mathfrak g\,,\quad
\left\{\begin{array}{ll}
	c=0 &\mbox{CYBE}\\
	c=\pm1 &\mbox{MCYBE}
\end{array}\right.\,.
\label{eq:YBE}
\end{equation}
In section \ref{sec:dilaton} we prove that the condition $K^a=0$ for the $\eta$-model is equivalent to the following algebraic condition on the $R$-matrix\footnote{Essentially the same condition was argued to appear from the analysis of vertex operators of the $\beta$-deformed $AdS_5\times S^5$ superstring in \cite{Bedoya:2010qz}, see equation $(87)$ there. That discussion would correspond to the truncation of our deformed action at order $\mathcal{O}(\eta^2)$. We thank Arkady Tseytlin for pointing this reference out to us.}
\begin{equation}
\mathrm{STr}(R\mathrm{ad}_x)=0\,,\qquad\forall x\in\mathfrak{g}\qquad(\mbox{i.e.}\qquad R^B{}_Af^A{}_{BC}=0)\,.
\label{eq:uni-R}
\end{equation}
We will refer to $R$-matrices satisfying this condition\footnote{It is easy to see that this condition is compatible with the (M)CYBE.} as ``unimodular'', for reasons that will be clear in section~\ref{sec:R-matrices}. Therefore the $\eta$-model has an interpretation as a string sigma model precisely for the unimodular $R$-matrices.

Let us consider the $\eta$-deformation based on an $R$-matrix which is a non-split\footnote{For the split case ($c=-1$) there exist no solution for the compact subalgebra $\mathfrak{su}(4)\subset \mathfrak{psu}(2,2|4)$. It seems then not possible to have a split solution for the full superalgebra.} ($c=1$ in (\ref{eq:YBE})) solution of the MCYBE for the supercoset on $AdS_5\times S^5$ with superalgebra $\mathfrak{psu}(2,2|4)$, as in~\cite{Delduc:2013qra}. A standard choice is to take  $R$ that multiplies by $-i\ (+i)$ positive (negative) roots of the complexified algebra, and annihilates Cartan elements. Choices of different real forms of the superalgebra correspond to inequivalent $R$-matrices, but one can check that none of the examples considered so far~\cite{Delduc:2013qra,Arutyunov:2015qva,Delduc:2014kha,Hoare:2016ibq} are unimodular, which is consistent with the findings of~\cite{Arutyunov:2015qva,Hoare:2016ibq}. 
We are not aware of a complete classification of solutions of the MCYBE for $\mathfrak{psu}(2,2|4)$, which leaves open the possibility of having unimodular non-split $R$-matrices that would lead to genuine string deformations. We will not analyze this question further here.

As first pointed out in~\cite{Kawaguchi:2014qwa}, there is a rich set of solutions to the CYBE ($c=0$ in (\ref{eq:YBE})) which can be used to define an $\eta$-deformation of the supercoset.
These $R$-matrices can be divided into two classes: abelian and non-abelian. Writing the $R$-matrix as (sums over repeated indices are understood)
\begin{equation}
R=\frac12r^{ij}b_i\wedge b_j\,,\qquad(R(x)=r^{ij}b_i\mathrm{Str}(b_jx),\  x\in\mathfrak g),
\label{eq:R-exp}
\end{equation}
abelian $R$-matrices are the ones for which $[b_i,b_j]=0\quad\forall i,j$ while non-abelian ones have $[b_i,b_j]\neq0$ for some $i,j$. The unimodularity condition (\ref{eq:uni-R}) takes the form
\begin{equation}
r^{ij}[b_i,b_j]=0\,.
\label{eq:bb-cond}
\end{equation}
This is trivially satisfied by any abelian $R$-matrix, which is consistent with observations in the literature, see e.g~\cite{Kyono:2016jqy,Hoare:2016hwh,Orlando:2016qqu}. 
This is also in line with the expectation that abelian $R$-matrices always have an interpretation in terms of (commuting) TsT-transformations\footnote{TsT stands for T-duality -- shift -- T-duality \cite{Lunin:2005jy,Frolov:2005ty,Frolov:2005dj}. 
Here we use it in the most general possible sense, e.g. including non-compact and fermionic T-dualities.}~\cite{vanTongeren:2015soa}. For non-abelian $R$-matrices the unimodularity condition (\ref{eq:bb-cond}) is non-trivial, and it is interesting to find all the compatible ones.
In fact, as explained in section~\ref{sec:R-matrices} it rules out most of the $R$-matrices of the so-called Jordanian type, which is the only class considered in the literature so far~\cite{Kawaguchi:2014qwa,vanTongeren:2015soa,Kyono:2016jqy,Hoare:2016hwh,Orlando:2016qqu}. 
	
Here we will focus on the problem of classifying all $R$-matrices which satisfy the CYBE  on the bosonic subalgebra $\mathfrak{so}(2,4)\oplus\mathfrak{so}(6)\subset \mathfrak{psu}(2,2|4)$ and are unimodular. 
The question is non-trivial only for non-abelian $R$-matrices, which we classify by the rank. 
From (\ref{eq:bb-cond}), any unimodular $R$-matrix of rank two $R=a\wedge b$ must be abelian, i.e. $[a,b]=0$, so non-abelian unimodular $R$-matrices have at least rank four.  
In tables \ref{tab:R-matrices} and \ref{tab:R-matrices-inner} we write down all non-abelian rank four $R$-matrices for $\mathfrak{so}(2,4)$ (the second table gives the inequvalent ones from the point of view of the string sigma model), and in table \ref{tab:R-matrices-iso-inner} we provide the bosonic isometries and the number of supersymmetries that they preserve. These $R$-matrices are constructed in section \ref{sec:R-matrices}, where we also show that the only other possibility is rank six. 
The extension to $\mathfrak{so}(2,4)\oplus\mathfrak{so}(6)$ is essentially trivial as it turns out that they must be abelian\footnote{This includes  $R$-matrices mixing generators of AdS and S, e.g. as in the so-called dipole deformations of~\cite{Gursoy:2005cn}}. in $\mathfrak{so}(6)$. Therefore there are no new marginal deformations of the dual CFT.\footnote{This statement remains to be true also if we further allow the $R$-matrix to act non-trivially on supercharges: after imposing unimodularity, preservation of the $\mathfrak{so}(2,4)$ isometry, reality and CYBE, we find that the only possible $R$-matrices are abelian and they act just on $\mathfrak{so}(6)$.}
Notice that $R_6, R_{13}$ and $R_{15}$ can be embedded in $\mathfrak{so}(2,3)$ and can therefore be used to define deformations of $AdS_4$. To have non-abelian deformations of $AdS_3$, instead, one must involve also generators from the sphere.

Because abelian $R$-matrices seem to generate backgrounds which can be equivalently obtained by doing (commuting) TsT-transformations on the undeformed model, one might suspect that $\eta$-deformed strings always correspond to TsT-transformations. With the exception of the last three $R$-matrices our results appear to be consistent with this expectation, see section~\ref{sec:R-matrices} for a discussion.

\vspace{24pt}

The outline of the rest of the paper is as follows.
In section~\ref{sec:models} we first review the definitions of the deformed models, we introduce a notation that highlights their similarities, and prove that the local fermionic symmetries of both deformed models are of the standard Green-Schwarz form.
In section~\ref{sec:geom-lambda} we derive the target space supergeometry for the $\lambda$-model, and by comparing to the results of~\cite{Wulff:2016tju} we extract the corresponding background fields.
Section~\ref{sec:geom-eta} achieves the same goal for the $\eta$-model. Here we also show how the unimodularity condition for the $R$-matrix is derived.
In section~\ref{sec:R-matrices} we study this condition in detail. We discuss its compatibility with Jordanian $R$-matrices, and derive all rank-four non-abelian unimodular $R$-matrices for $\mathfrak{so}(2,4)$ which solve the CYBE.
In section~\ref{sec:backgrounds} we consider the case of backgrounds generated by $R$-matrices which act only on the bosonic subalgebra. We work out certain examples generated by the $R$-matrices previously derived, and we check in some cases that  they are equivalent to sequences of TsT transformations on the original undeformed model.

\vspace{48pt}

\begin{table}[h]
$$
\begin{array}{l}
R_1=p_1\wedge p_2+(p_0+p_3)\wedge(J_{01}-J_{13})\\
R_2=p_1\wedge p_2+(p_0+p_3)\wedge(p_3+J_{01}-J_{13})\\
R_3=p_1\wedge(J_{02}-J_{23})+(p_0+p_3)\wedge(p_2+J_{01}-J_{13})\\
R_4=(p_1-J_{02}+J_{23})\wedge(k_0+k_3+2p_3-2J_{12})+2(p_0+p_3)\wedge(p_2+J_{01}-J_{13})\\
R_5=p_1\wedge(J_{02}-J_{23})+(p_0+p_3)\wedge(D+J_{03})\\
R_6=p_1\wedge J_{03}+2p_0\wedge p_3\\
R_7=J_{03}\wedge J_{12}+2p_0\wedge p_3\\
R_8=p_1\wedge p_2+(p_0+p_3)\wedge J_{12}\\
R_9=p_1\wedge p_2+(p_0+p_3)\wedge(p_3+J_{12})\\
R_{10}=p_1\wedge p_2+p_3\wedge(p_0+J_{12})\\
R_{11}=p_1\wedge p_2+p_3\wedge J_{12}\\
R_{12}=p_1\wedge p_2+p_0\wedge(p_3+J_{12})\\
R_{13}=p_1\wedge p_2+p_0\wedge J_{12}\\
R_{14}=p_1\wedge p_2+J_{12}\wedge J_{03}\\
R_{15}=p_1\wedge p_3+(J_{01}-J_{13})\wedge(p_0+p_3)\\
R_{16}=p_1\wedge p_3+(p_2+J_{01}-J_{13})\wedge(p_0+p_3)\\
R_{17}=p_1\wedge(p_3+J_{02}-J_{23})+(p_0+p_3)\wedge(p_2+J_{01}-J_{13})
\end{array}
$$
\caption{All non-abelian unimodular rank-four $R$-matrices (CYBE) of $\mathfrak{so}(2,4)$ up to automorphisms of the corresponding subalgebras (see section \ref{sec:R-matrices}).
}
\label{tab:R-matrices}
\end{table}


\begin{table}[t]
$$
\begin{array}{l}
R_1=(p_1+a (J_{01}-J_{13}))\wedge p_2+(p_0+p_3)\wedge(J_{01}-J_{13})\\   %
R_2=(p_1+a(p_3+J_{01}-J_{13})+b(p_0+p_3)) \wedge p_2+(p_0+p_3)\wedge(p_3+J_{01}-J_{13})\\  %
R_3=(p_1+a(p_2+J_{01}-J_{13}))\wedge(p_1+J_{02}-J_{23})+(p_0+p_3)\wedge(p_2+J_{01}-J_{13})\\  %
R_4=((p_1-J_{02}+J_{23})+2a(p_2+J_{01}-J_{13})+2b(p_0+p_3))\wedge(k_0+k_3+2p_3-2J_{12}+c(p_0+p_3))\\ %
\qquad+2d(p_0+p_3)\wedge(p_2+J_{01}-J_{13})\\
R_5=p_1\wedge(J_{02}-J_{23})+a(p_0+p_3)\wedge(D+J_{03})\\   %
R_6=p_1\wedge J_{03}+2p_0\wedge p_3\\ %
R_7=J_{03}\wedge J_{12}+2p_0\wedge p_3\\ %
R_8=p_1\wedge p_2+(p_0+p_3)\wedge J_{12}\\ %
R_9=p_1\wedge p_2+a\, (p_0+p_3)\wedge(p_3+J_{12})\\ %
R_{10}=p_1\wedge p_2+a\, p_3\wedge(p_0+J_{12})\\  %
R_{11}=p_1\wedge p_2+ p_3\wedge J_{12}\\ %
R_{12}=p_1\wedge p_2+a\, p_0\wedge(p_3+J_{12})\\ %
R_{13}=p_1\wedge p_2+ p_0\wedge J_{12}\\ %
R_{14}=p_1\wedge p_2+ J_{12}\wedge J_{03}\\ %
R_{15}=(p_1+a(p_0+p_3))\wedge p_3+(J_{01}-J_{13})\wedge(p_0+p_3)\\ %
R_{16}=(p_1+a(p_0+p_3))\wedge p_3+(p_2+J_{01}-J_{13})\wedge(p_0+p_3)\\
R_{17}=(p_1+a(p_0+p_3))\wedge(p_1+p_3+J_{02}-J_{23})+(p_0+p_3)\wedge(p_2+J_{01}-J_{13})
\end{array}
$$
\caption{All non-abelian unimodular rank four $R$-matrices (CYBE) of $\mathfrak{so}(2,4)$ up to inner automorphisms.}
\label{tab:R-matrices-inner}
\end{table}
\begin{table}[h!]
$$
\begin{array}{l|c|l}
& \text{supercharges} & \text{bosonic isometries}\\
\hline
R_1 & 8 & p_0+p_3,\ p_1 ,\ p_2 ,\ p_0 -p_3-2(J_{02} - J_{23}) ,\quad  (a= 0) \\
 & 8 & p_0+p_3,\ p_1 +a (J_{01} - J_{13}),\ p_2 ,\ \qquad\qquad\ \ (a\neq 0) \\
R_2 & 8 & p_0+p_3,\  p_1 ,\ p_2 ,\ p_0-  p_3-J_{01} - J_{02} + J_{13} + J_{23} ,\quad \ (a=0)\\
 & 8 & p_0+p_3,\  p_1 +a (J_{01} - J_{13}) ,\ p_2 ,\qquad\qquad\qquad\qquad\quad  (a\neq 0)\\
R_3 & 8 &  p_0 +p_3,\ p_1 ,\ J_{02} - J_{23} ,\ \qquad\qquad\qquad\qquad\qquad\qquad\qquad\quad\ (a=0)\\
 & 8 &  p_0 +p_3,\  p_1 + (J_{02} - J_{23}) ,\ J_{02} - J_{23} - a (J_{01} - J_{13} + p_2) ,\quad  (a\neq 0)\\
R_4 & 0 &  -J_{02} + J_{23} + p_1 +  2 a (J_{01} - J_{13} + p_2),\   p_0+p_3,\ 2 J_{12} -2 p_3 -k_0 - k_3,\ \\
R_5 & 8 & D + J_{03} ,\ p_0 +p_3,\ \\
R_6 & 0 & J_{03} ,\ p_1,\  p_2, \ \\
R_7 & 0 & J_{03},\ J_{12}, \\
R_8 & 0 &  p_0, \ p_3,\ J_{12}, \\
R_9 & 0 &  p_0, \ p_3,\ J_{12}, \\
R_{10} & 0 &  p_0, \ p_3,\ J_{12}, \\
R_{11} & 0 &  p_0, \ p_3,\ J_{12}, \\
R_{12} & 0 &  p_0, \ p_3,\ J_{12}, \\
R_{13} & 0 &  p_0, \ p_3,\ J_{12}, \\
R_{14} & 0 & J_{03},\ J_{12}, \\
R_{15} & 8 & p_0+p_3,\ p_1,\ p_2,\ \\
R_{16} & 8 & p_0+p_3,\ p_1,\ p_2,\ \\
R_{17} & 8 & p_0+p_3 ,\  p_1,\ J_{02} - J_{23} -p_2 + p_3,
\end{array}
$$
\caption{For each $R$-matrix of Table~\ref{tab:R-matrices-inner} we indicate the number of unbroken supercharges and we list the unbroken bosonic isometries.}
\label{tab:R-matrices-iso-inner}
\end{table}

\newpage

\section{$\eta$ and $\lambda$-deformed string sigma models}\label{sec:models}
The $\eta$ and $\lambda$ deformations are deformations of supercoset sigma models that preserve the classical integrability of the original models. In the string theory context the most studied example is the deformation of the $AdS_5\times S^5$ string\footnote{Another supercoset closely related to this is the pp-wave background of~\cite{Metsaev:2001bj}.} described by a $\frac{PSU(2,2|4)}{SO(1,4)\times SO(5)}$ supercoset sigma model~\cite{Metsaev:1998it}. However, there are many other backgrounds where at least a subsector of the string worldsheet theory is described by a supercoset sigma model, e.g. $AdS_4\times\mathbbm{CP}^3$~\cite{Arutyunov:2008if,Stefanski:2008ik,Gomis:2008jt}, $AdS_3\times S^3\times T^4$~\cite{Babichenko:2009dk}, $AdS_2\times S^2\times T^6$~\cite{Sorokin:2011rr} and several others~\cite{Wulff:2014kja}. 

We start by reviewing the definitions of the deformed models.
The relevant superalgebra conventions are collected in appendix \ref{app:superalgebras}.

\subsection{Lagrangians of the deformed models}
The $\eta$-model Lagrangian takes the form~\cite{Delduc:2013qra,Kawaguchi:2014qwa}
\begin{equation}
\mathcal L=-\frac{(1+c\eta^2)^2}{4(1-c\eta^2)}(\gamma^{ij}-\varepsilon^{ij})\mathrm{Str}(g^{-1}\partial_ig\ \hat d\, \mathcal O_-^{-1}(g^{-1}\partial_jg))\,,
\label{eq:L-eta}
\end{equation}
where $g$ is a group element of $G$, $i,j$ are worldsheet indices, $\gamma^{ij}$ is the (Weyl-invariant) worldsheet metric and $\varepsilon^{01}=+1$.
Here $\eta$ is the deformation parameter, and setting $\eta=0$ yields the Lagrangian of the undeformed supercoset sigma model. The deformation involves the Lie algebra operators
\begin{equation}\label{eq:Opm-eta}
\mathcal O_+=1+\eta R_g\hat d^T\,,\qquad \mathcal O_-=1-\eta R_g\hat d\,,
\end{equation}
where $R_g=\Ad_g^{-1}R\Ad_g$, $R^T=-R$ and $R$ satisfies the (M)CYBE (\ref{eq:YBE}). 
Our derivation is general and we will not need to pick a particular solution of (\ref{eq:YBE}): we only need to assume the above properties for $R$, and we will treat the homogeneous ($c=0$, CYBE) and the inhomogeneous ($c=1$, MCYBE) cases at the same time. 
In the Lagrangian the following combinations of projection operators appear
\begin{equation}\label{eq:dplusdT}
\begin{aligned}
\hat d&=P^{(1)}+2\hat\eta^{-2}P^{(2)}-P^{(3)}\,,\qquad \qquad&&\hat\eta=\sqrt{1-c\eta^2}\,.\\
\hat d^T&=-P^{(1)}+2\hat\eta^{-2}P^{(2)}+P^{(3)}\,, \qquad &&\text{where }\hat d+\hat d^T=4\hat\eta^{-2}P^{(2)}\,.
\end{aligned}
\end{equation}
The $\lambda$-model is defined as a deformation of the $G/G$ gauged WZW model. To get a standard string sigma model one integrates out the gauge-field which leads to a Lagrangian\footnote{This is the classical Lagrangian. At the quantum level there is also a Fradkin-Tseytlin term $R^{(2)}\phi$ present, where $\phi$ is the dilaton superfield, generated by integrating out the gauge-field, whose form will be discussed in section \ref{sec:geom-lambda}.} somewhat similar to that of the $\eta$-model, namely~\cite{Hollowood:2014qma}
\begin{equation}
\mathcal L=-\frac{k}{2\pi}(\gamma^{ij}-\varepsilon^{ij})\mathrm{Str}(g^{-1}\partial_ig(1+\widehat B_0-2\mathcal O_-^{-1})(g^{-1}\partial_jg))\,.
\label{eq:L-lambda}
\end{equation}
Here $k$ is the level of the WZW model,\footnote{$\widehat B_0=-\widehat B_0^T$ is related to the original WZ-term, see section \ref{sec:geom-lambda}.} and the Lie algebra operators $\mathcal O_\pm$ are now defined as
\begin{equation}
\mathcal O_+=\Ad_g^{-1}-\Omega^T\,,\qquad \mathcal O_-=1-\Ad_g^{-1}\Omega\,.
\label{eq:Opm-lambda}
\end{equation}
In this case things are written in terms of the combinations of projectors
\begin{equation}\label{eq:omegaomegaT}
\begin{aligned}
\Omega&=P^{(0)}+\lambda^{-1}P^{(1)}+\lambda^{-2}P^{(2)}+\lambda P^{(3)}\,,&& \\
\Omega^T&=P^{(0)}+\lambda P^{(1)}+\lambda^{-2}P^{(2)}+\lambda^{-1}P^{(3)}\,, \qquad 
&&1-\Omega\Omega^T=1-\Omega^T\Omega=(1-\lambda^{-4})P^{(2)}\,.
\end{aligned}
\end{equation}

Both the Lagrangian (\ref{eq:L-eta}) of the $\eta$ and (\ref{eq:L-lambda}) of the $\lambda$-model can be formally written in the same way\footnote{We have used (\ref{eq:dplusdT}), (\ref{eq:omegaomegaT}), $\Ad_g^T=\Ad_g^{-1}$ and $R_g^T=-R_g$.}
\begin{equation}
\mathcal L=-\frac{T}{2}\gamma^{ij}\mathrm{Str}(A_{-i}^{(2)}A_{-j}^{(2)})+\frac{T}{2}\varepsilon^{ij}\mathrm{Str}(A_{-i}\widehat BA_{-j})\,,
\label{eq:L-eta-lambda}
\end{equation}
in terms of the one-forms
\begin{equation}
A_\pm=\mathcal O_\pm^{-1}(g^{-1}dg)\,,
\end{equation}
where the string tension $T$ and the operator $\widehat B$ (responsible for the $B$-field) in the two cases are
\begin{equation}
\begin{split}
\mathbf\eta-\mbox{\bf model}:&\qquad T=\left(\frac{1+c\eta^2}{1-c\eta^2}\right)^2\,,\qquad \widehat B=\frac{\hat\eta^2}{2}(P^{(1)}-P^{(3)}+\eta\hat d^T R_g\hat d)\,,
\label{eq:B-field}
\\
\mathbf\lambda-\mbox{\bf model}:&\qquad T=\frac{k}{\pi}(\lambda^{-4}-1)\,,\qquad\widehat B=(\lambda^{-4}-1)^{-1}(\mathcal O_-^T\widehat B_0\mathcal O_-+\Omega^T\Ad_g-\Ad_g^{-1}\Omega)\,.
\end{split}
\end{equation}
An important role is played by the operator
\begin{equation}\label{eq:def-M}
M=\mathcal O_-^{-1}\mathcal O_+
\end{equation}
which relates $A_-$ to $A_+$ as $A_-=MA_+$.
Using the expressions in~\eqref{eq:M} it is not hard to show that
\begin{equation}
M^TP^{(2)}M=P^{(2)}\,,
\end{equation}
which implies that the operator $P^{(2)}MP^{(2)}$ implements a Lorentz transformation on the subspace with grading-2 of the superisometry algebra. This implies that there exists an element $h\in H=G^{(0)}\subset G$ such that
\begin{equation}
P^{(2)}MP^{(2)}=\Ad_h^{-1}P^{(2)}=P^{(2)}\Ad_h^{-1}\,.
\label{eq:adh}
\end{equation}
The fact that $\Ad_h$ is a Lorentz transformation implies the basic relation between the action on vectors and spinors
\begin{equation}
[\Ad_h]^{\hat\gamma}{}_{\hat\alpha}\gamma^a_{\hat\gamma\hat\delta}[\Ad_h]^{\hat\delta}{}_{\hat\beta}=[\Ad_h]^a{}_b\gamma^b_{\hat\alpha\hat\beta}\,.
\end{equation}
We refer to appendix~\ref{app:def} for some useful identities satisfied by the operators entering the deformed models.

\subsection{Kappa symmetry transformations in Green-Schwarz form}\label{sec:kappa}
Both the $\eta$ and $\lambda$ model have a local fermionic symmetry which removes 16 of the 32 fermions, and here we show that it takes the form of the standard kappa symmetry of the GS superstring. The transformations for the local fermionic symmetry take the form~\cite{Delduc:2013qra,Kawaguchi:2014qwa,Hollowood:2014qma}
\begin{equation}
\mathcal O_+^{-1}(g^{-1}\delta_\kappa g)=P_-^{ij}\{i\tilde\kappa_i^{(1)},A_{-j}^{(2)}\}+\zeta^sP_+^{ij}\{i\tilde\kappa_i^{(3)},A_{+j}^{(2)}\}\,,
\end{equation}
where we denote the parameter by $\tilde\kappa$, which is related to the kappa symmetry parameter $\kappa$ of the GS string as explained below.
The above transformations are accompanied by the variation of the worldsheet metric
\begin{equation}
\delta_\kappa\gamma^{ij}=
\frac{\zeta^2}{2}\big(\mathrm{Str}(W[(P_+i\tilde\kappa^{(1)})^i,{(P_+A_+^{(1)})}^j])+\mathrm{Str}(W[(P_-i\tilde\kappa^{(3)})^i,{(P_-A_-^{(3)})}^j])\big)
\,,
\end{equation}
where we have defined
\begin{equation}
P_\pm^{ij}=\frac12(\gamma^{ij}{\pm}\varepsilon^{ij})
\,,\qquad
\zeta=
\left\{
\begin{array}{c}
\hat\eta\\
\lambda
\end{array}
\right.
\,,\qquad
s=
\left\{
\begin{array}{c}
0 \qquad \mathbf\eta-\mbox{\bf model}\\
1 \qquad \mathbf\lambda-\mbox{\bf model}
\end{array}
\right.
\,.
\end{equation}
Using the fact that $A^{(2)}_-$ is related to $A_+^{(2)}$ by a gauge transformation, i.e.
\begin{equation}
A_-^{(2)}=P^{(2)}MA_+^{(2)}=\Ad_h^{-1}A_+^{(2)}\,,
\end{equation}
we can write the kappa transformations as\footnote{In writing the transformations in this form we used (\ref{eq:Apm-rel}).}
\begin{equation}
\begin{aligned}
&i_{\delta_\kappa}E^{(2)}=0\,,\qquad
i_{\delta_\kappa}E^{(1)}=P_-^{ij}\{i\kappa_i^{(1)},E_j^{(2)}\}\,,\qquad
i_{\delta_\kappa}E^{(3)}=P_+^{ij}\{i\kappa_i^{(3)},E_j^{(2)}\}
\\
&\delta_\kappa\gamma^{ij}=
\frac12\mathrm{Str}(W[(P_+i\kappa^{(1)})^i,{(P_+E^{(1)})}^j])
+\frac12\mathrm{Str}(W[(P_-i\kappa^{(3)})^i,{(P_-E^{(3)})}^j])\,,
\end{aligned}
\end{equation}
where $\kappa^{(1)}=\zeta\Ad_h\tilde\kappa^{(1)}$ and $\kappa^{(3)}=(-i)^s\zeta\tilde\kappa^{(3)}$. This shows that the kappa symmetry variations have the standard GS form, and at the same time it allows us to identify the supervielbeins with projections of $A_\pm$ as\footnote{The explicit $i$ in $E^{(3)}$ and $\kappa^{(3)}$ in the case of the $\lambda$-model is needed to put the transformations in the standard type IIB form. The reason for having $i$ can be traced to the relative sign between $P^{(1)}$ and $P^{(3)}$ in (\ref{eq:omegaomegaT}) compared to (\ref{eq:dplusdT}). Alternatively, insisting on manifest reality of the model, the kappa symmetry transformations and superspace constraints become those of type IIB* rather than type IIB. This is rather natural since the $\lambda$-model is a deformation of the non-abelian T-dual of the $AdS_5\times S^5$ string, which involves also a T-duality in the time direction.}
\begin{equation}
E^{(2)}\equiv E^aP_a=A_+^{(2)}\,,\qquad
E^{(1)}\equiv E^{\hat\alpha1}Q^1_{\hat\alpha}=\zeta\Ad_hA_+^{(1)}\,,\qquad
E^{(3)}\equiv E^{\hat\alpha2}Q^2_{\hat\alpha}=i^s\zeta A_-^{(3)}\,.
\label{eq:supervielbeins}
\end{equation}
In terms of these the Lagrangian (\ref{eq:L-eta-lambda}) takes the standard form
\begin{equation}
L=-\frac{T}{2}\gamma^{ij}\mathrm{Str}(E_i^{(2)}E_j^{(2)})
+\frac{T}{2}\varepsilon^{ij}B_{ij}\,,
\end{equation}
where the $B$-field can be read off from (\ref{eq:B-field}).

Since the action and kappa symmetry transformations take the standard GS form, it follows from the analysis of~\cite{Wulff:2016tju} that the target superspace of these models solves the generalized type II supergravity equations derived there. If the Killing vector $K^a$ appearing in these equations vanishes, they reduce to the standard supergravity equations. 
In the next sections we will derive the form of the target space supergeometry for the $\eta$ and $\lambda$-deformed strings. 
Having identified the supervielbeins of the background superspace we can find the supergeometry by calculating the torsion\footnote{Our conventions are the same as those of~\cite{Wulff:2016tju}. In particular $d$ acts from the right and components of superforms are defined as $\omega_n=\frac{1}{n!}E^{A_n}\wedge\cdots\wedge E^{A_1}\omega_{A_1\cdots A_n}$.}
\begin{equation}
T^a=dE^a+E^b\wedge\Omega_b{}^a\,,\qquad T^{\hat\alpha I}=dE^{\hat\alpha I}-\frac14(\gamma_{ab}E^I)^{\hat\alpha}\wedge\Omega^{ab}\qquad(I=1,2)\,,
\label{eq:torsion}
\end{equation}
and reading off the background superfields by comparing to the general expressions derived in~\cite{Wulff:2016tju}. These are valid for a generalized type II supergravity background and reduce to those of a standard supergravity background (see e.g.~\cite{Wulff:2013kga}) only when $K^a=0$.
We will see that the $\lambda$-model background is a solution to standard (type II*) supergravity. For the $\eta$-model background we will derive the condition on the $R$-matrix of the $\eta$-model for it to give rise to a standard type II background.

\section{Target superspace for the $\lambda$-model}\label{sec:geom-lambda}
In this section we present the derivation for the $\lambda$-model. We refer to appendix~\ref{app:lambda} for more details.
The supervielbeins are defined in terms of projections of $A_\pm$ by (\ref{eq:supervielbeins}). To calculate the torsion we therefore need to calculate the exterior derivative of $A_\pm$. Using $A_+=\mathcal O^{-1}_+(g^{-1}dg)$ where $\mathcal O_\pm$ are defined in (\ref{eq:Opm-lambda}) we find 
\begin{align}
dA_+=&
\mathcal O_+^{-1}(d\mathcal O_+\wedge A_+)
+\mathcal O_+^{-1}(g^{-1}dg\wedge g^{-1}dg)
\nonumber\\
=&
-\mathcal O_+^{-1}\{g^{-1}dg,\Ad_g^{-1}A_+\}
+\frac12\mathcal O_+^{-1}\{g^{-1}dg,g^{-1}dg\}
\nonumber\\
=&
-\frac12\mathcal O_+^{-1}\{\Ad_g^{-1}A_+,\Ad_g^{-1}A_+\}
+\frac12\mathcal O_+^{-1}\{\Omega^TA_+,\Omega^TA_+\}
\nonumber\\
=&
-\frac12\{A_+,A_+\}
-\frac12\mathcal O_+^{-1}(\Omega^T\{A_+,A_+\}-\{\Omega^TA_+,\Omega^TA_+\})\,,
\label{eq:dAp-lambda}
\end{align}
where we used the fact that $g^{-1}dg=\mathcal O_+A_+=(\Ad_g^{-1}-\Omega^T)A_+$ to write everything in terms of $A_+$. An almost identical calculation gives
\begin{equation}\label{eq:dAm-lambda}
dA_-=
\frac12\{A_-,A_-\}
+\frac12\mathcal O_-^{-1}\Ad_g^{-1}(\Omega\{A_-,A_-\}-\{\Omega A_-,\Omega A_-\})\,.
\end{equation}
In the above equations it is useful to expand out the expressions inside parenthesis, see~\eqref{eq:dA-plus-lambda},~\eqref{eq:dA-minus-lambda}. Projecting equation~\eqref{eq:dA-plus-lambda} with $P^{(2)}$ we find
\begin{align}
dE^{(2)}=&
\frac12\{E^{(1)},E^{(1)}\}
+\frac12\{E^{(3)},E^{(3)}\}
-\{A_+^{(0)},E^{(2)}\}
-i\lambda\{E^{(3)},P^{(3)}ME^{(2)}\}
\nonumber\\
&{}
-i\lambda P^{(2)}M^T\{E^{(2)},E^{(3)}\}
-\frac12\lambda^2\{P^{(3)}ME^{(2)},P^{(3)}ME^{(2)}\}
-\frac12P^{(2)}M^T\{E^{(2)},E^{(2)}\}
\nonumber\\
&{}
-\lambda^2P^{(2)}M^T\{E^{(2)},P^{(3)}ME^{(2)}\}\,.
\label{eq:dE2-lambda}
\end{align}
where the result has been rewritten in terms of the supervielbeins (\ref{eq:supervielbeins}), and we have used (\ref{eq:Apm-rel}) and (\ref{eq:adh}).
Using the explicit form of the commutators in (\ref{eq:alg-b}) and (\ref{eq:alg-f}) we find that the component $T^a$ of the torsion takes the standard form (here and in the following we drop the $\wedge$'s for readability)
\begin{align}
T^a
=
dE^a
+E^b\Omega_b{}^a
=
-\frac{i}{2}E^1\gamma^aE^1
-\frac{i}{2}E^2\gamma^aE^2
\,,
\label{eq:bos-torsion}
\end{align}
if we identify the spin connection as\footnote{Here we rewrote $A_\pm^{(0)}=\frac12A_\pm^{ab}J_{ab}$ and used the relation between components of $M$ and $M^T$ in (\ref{eq:MT-comp}).}
\begin{equation}
\Omega_{ab}
=
-(A_+)_{ab}
-2\lambda(E^2\gamma_{[a})_{\hat\alpha}M^{\hat\alpha2}{}_{b]}
-\frac{3i}{2}\lambda^2E^cM^{\hat\alpha2}{}_{[a}(\gamma_b)_{\hat\alpha\hat\beta}M^{\hat\beta2}{}_{c]}
+\frac12E^c(M_{ab,c}-2M_{c[a,b]})\,.
\label{eq:connection-lambda}
\end{equation}

To derive the other components of the torsion we first need to compute the exterior derivative of the fermionic supervielbeins.
Using (\ref{eq:dA-minus-lambda}) and (\ref{eq:supervielbeins}) we find
\begin{align}
dE^{(3)}=&
\frac{i}{2}\lambda P^{(3)}M\{E^{(3)},E^{(3)}\}
-\{A_+^{(0)},E^{(3)}\}
+\{P^{(0)}ME^{(2)},E^{(3)}\}
\nonumber\\
&{}
-i\lambda\big[1+\lambda(1-\lambda^{-4})P^{(3)}(\mathcal O_+^T)^{-1}\big]\Ad_h^{-1}\big(\{E^{(2)},E^{(1)}\}-\{E^{(2)},\Ad_hP^{(1)}ME^{(2)}\}\big)
\nonumber\\
&{}
+\frac{i}{2}\lambda(1-\lambda^{-4})P^{(3)}(\mathcal O_+^T)^{-1}\Ad_h^{-1}\{E^{(2)},E^{(2)}\}\,.
\end{align}
Since we have already identified the form of the spin connection (\ref{eq:connection-lambda}) from the previous computation, we can now find the corresponding component of the torsion (\ref{eq:torsion}) and compare it to the standard form given in~\cite{Wulff:2016tju}, i.e.
\begin{equation}\label{eq:ferm2-torsion}
T^{\hat\alpha2}
=
E^{\hat\alpha2}\,E^2\chi^2
-\frac12E^2\gamma^aE^2(\gamma_a\chi^2)^{\hat\alpha}
+\frac18E^a(E^2\gamma^{bc})^{\hat\alpha}H_{abc}
-\frac18E^a(E^1\gamma_a\mathcal S^{12})^{\hat\alpha}
+\frac12E^bE^a\psi_{ab}^{\hat\alpha2}\,,
\end{equation}
where $H$ is the NSNS three-form, $\mathcal S$ the RR bispinor, $\chi^I_{\hat\alpha}$ the dilatino and $\psi_{ab}^{\hat\alpha I}$ the gravitino field strength superfields.
We find that $T^{\hat\alpha2}$ takes the above form if we identify 
\begin{align}
H_{abc}=&3M_{[ab,c]}+3i\lambda^2M^{\hat\alpha2}{}_{[a}(\gamma_b)_{\hat\alpha\hat\beta}M^{\hat\beta2}{}_{c]}\,,
\label{eq:Habc-lambda}
\\
\mathcal S^{\hat\alpha1\hat\beta2}=&
-8\lambda\big[\Ad_h(1+\lambda(1-\lambda^{-4})\mathcal O_+^{-1})\big]^{\hat\alpha1}{}_{\hat\gamma1}\widehat{\mathcal K}^{\hat\gamma1\hat\beta2}\,,
\label{eq:S-lambda}
\\
\chi^2_{\hat\alpha}=&\frac12\lambda\gamma^a_{\hat\alpha\hat\beta}M^{\hat\beta2}{}_a\,,
\label{eq:chi2-lambda}
\\
\psi_{ab}^{\hat\alpha2}=&
\frac{i}{4}\lambda(1-\lambda^{-4})[(\mathcal O_+^T)^{-1}\Ad_h^{-1}]^{\hat\alpha2}{}_{cd}\widehat{\mathcal K}_{ab}{}^{cd}
-\frac14[\Ad_hM]^{\hat\beta1}{}_{[a}(\gamma_{b]})_{\hat\beta\hat\gamma}\mathcal S^{\hat\gamma1\hat\alpha2}\,.
\end{align}
As already remarked, the RR bispinor superfield is imaginary if we interpret the $\lambda$-model target space as a solution of type II supergravity, as here, rather than type II* supergravity.\footnote{Let us recall that at least in some cases it is possible to define a real type II background, after analytic continuation or proper choice of coordinate patch~\cite{Borsato:2016zcf,Chervonyi:2016ajp}.} This determines the bosonic target space fields, with the exception of the dilaton which we will determine shortly. First, let us calculate also the remaining components of the femionic superfields, which we will extract from the corresponding component of the torsion, $T^{\hat\alpha1}$. From (\ref{eq:dA-plus-lambda}) and using (\ref{eq:supervielbeins}) we find
\begin{align}
dE^{(1)}
=&
-\{\Ad_hA_+^{(0)}+dhh^{-1},E^{(1)}\}
+\frac12\lambda(1-\lambda^{-4})P^{(1)}\Ad_h\mathcal O_+^{-1}\Ad_h^{-1}\{E^{(1)},E^{(1)}\}
\nonumber\\
{}&
-i\lambda\Ad_h\{E^{(2)},E^{(3)}\}
-\lambda^2\Ad_h\{E^{(2)},P^{(3)}ME^{(2)}\}
-i\lambda^2(1-\lambda^{-4})P^{(1)}\Ad_h\mathcal O_+^{-1}\{E^{(2)},E^{(3)}\}
\nonumber\\
{}&
-\frac12\lambda(1-\lambda^{-4})P^{(1)}\Ad_h\mathcal O_+^{-1}
\big(
\{E^{(2)},E^{(2)}\}
+2\lambda^2\{E^{(2)},P^{(3)}ME^{(2)}\}
\big)\,.
\label{eq:dE1-lambda}
\end{align}
Using this expression we find\footnote{To calculate this component of the torsion we must first find the Lorentz-transformed spin connection $\Ad_hA_+^{(0)}+dhh^{-1}$ appearing in the first term, see equation~\eqref{eq:dh-lambda} and the corresponding derivation.}
\begin{equation}\label{eq:ferm1-torsion}
T^{\hat\alpha1}
=
E^{\hat\alpha1}\,E^1\chi^1
-\frac12E^1\gamma^aE^1(\gamma_a\chi^1)^{\hat\alpha}
-\frac18E^a(E^1\gamma^{bc})^{\hat\alpha}H_{abc}
-\frac18E^a(E^2\gamma_a\mathcal S^{21})^{\hat\alpha}
+\frac12E^bE^a\psi_{ab}^{\hat\alpha1}\,,
\end{equation}
is again of the standard form given in~\cite{Wulff:2016tju}, where $\mathcal S^{\hat\beta2\hat\alpha1}=-\mathcal S^{\hat\alpha1\hat\beta2}$ and
\begin{align}
\chi^1_{\hat\alpha}=-\frac{i}{2}\gamma^b_{\hat\alpha\hat\beta}[\Ad_hM]^{\hat\beta1}{}_b\,,
\label{eq:chi1-lambda}
\qquad
\psi_{ab}^{\hat\alpha1}=
-\frac12\lambda(1-\lambda^{-4})[\Ad_h\mathcal O_+^{-1}]^{\hat\alpha1}{}_{cd}\widehat{\mathcal K}_{ab}{}^{cd}
-\frac{i}{4}\lambda(\mathcal S^{12}\gamma_{[a})^{\hat\alpha}{}_{\hat\beta} M^{\hat\beta2}{}_{b]}\,.
\end{align}
We complete the set of background superfields for the $\lambda$-model by noting that the $B$-field can be written in the two equivalent forms
\begin{equation}\label{eq:B-lambda}
\begin{aligned}
B&=(\lambda^{-4}-1)^{-1}\big[B_0+\mathrm{Str}(g^{-1}dg\wedge A_-)\big]\,, && \quad dB_0=\frac13\mathrm{Str}(g^{-1}dg\wedge g^{-1}dg\wedge g^{-1}dg)\,,\\
&=(\lambda^{-4}-1)^{-1}\big[B_0-\mathrm{Str}(g^{-1}dg\wedge\Omega^TA_+)\big]\,, &&
\end{aligned}
\end{equation}
and that the dilaton is given by
\begin{equation}\label{eq:e-2phi-lambda}
e^{-2\phi}=\mathrm{sdet}(\mathcal O_+)=\mathrm{sdet}(\Ad_g-\Omega)\,.
\end{equation}
This result for the dilaton arises from integrating out the gauge-fields in the deformed gauged WZW model~\cite{Hollowood:2014qma}. 
To verify that the $\lambda$-model gives rise to a standard supergravity background\footnote{As pointed out in~\cite{Wulff:2016tju} this was clear from the fact that the metric of the $\lambda$-model does not admit any isometries, so that the Killing vector $K^a$ of the generalized supergravity equations vanishes.} it is enough to verify that the dilatino's found in (\ref{eq:chi2-lambda}) and (\ref{eq:chi1-lambda}) are indeed the spinor derivatives of $\phi$
\begin{equation}
\begin{aligned}
\nabla_{\hat\alpha2}\phi&
=
\frac{i}{2}\lambda\widehat{\mathcal K}^{\hat\beta1\hat\gamma2}\mathrm{STr}(Q^1_{\hat\beta} M[Q^2_{\hat\alpha},Q^2_{\hat\gamma}])
=
\chi^2_{\hat\alpha}\,,
\\
\nabla_{\hat\alpha1}\phi&
=
\frac12(1-\lambda^{-4})[\Ad_h^{-1}]^{\hat\beta}{}_{\hat\alpha}\mathrm{STr}(P^a\mathcal O_-^{-1}[Q^1_{\hat\beta},P_a])
=
\chi^1_{\hat\alpha}\,.
\end{aligned}
\end{equation}

\section{Target superspace for the $\eta$-model}\label{sec:geom-eta}
The calculations for the $\eta$-model proceed along the same lines as those for the $\lambda$-model with only minor differences. We begin by calculating the derivative of $A_+$
\begin{align}
dA_+=\,&
\mathcal O_+^{-1}(d\mathcal O_+\wedge A_+)
+\mathcal O_+^{-1}(g^{-1}dg\wedge g^{-1}dg)
\nonumber\\
=\,&
\eta\mathcal O_+^{-1}R_g\{g^{-1}dg,\hat d^TA_+\}
-\eta\mathcal O_+^{-1}\{g^{-1}dg,R_g\hat d^TA_+\}
+\frac12\mathcal O_+^{-1}\{g^{-1}dg,g^{-1}dg\}
\nonumber\\
=\,&
\frac12\mathcal O_+^{-1}\{A_+,A_+\}
+\eta\mathcal O_+^{-1}R_g\{A_+,\hat d^TA_+\}
+\eta^2\mathcal O_+^{-1}R_g\{R_g\hat d^TA_+,\hat d^TA_+\}
\nonumber\\
&{}
-\frac12\eta^2\mathcal O_+^{-1}\{R_g\hat d^TA_+,R_g\hat d^TA_+\}
\nonumber\\
=\,&
\frac12\mathcal O_+^{-1}\{A_+,A_+\}
-\frac12c\eta^2\mathcal O_+^{-1}\{\hat d^TA_+,\hat d^TA_+\}
+\eta\mathcal O_+^{-1}R_g\{A_+,\hat d^TA_+\}\,,
\label{eq:dAp-eta}
\end{align}
where we used the fact that $g^{-1}dg=\mathcal O_+A_+$ and in the last step we used the fact that $R$ (as well as $R_g$) satisfies the (M)CYBE equation, so that
\begin{equation}
\{R_g\hat d^TA_+,R_g\hat d^TA_+\}
-2R_g\{R_g\hat d^TA_+,\hat d^TA_+\}
-c\{\hat d^TA_+,\hat d^TA_+\}
=0\,.
\end{equation}
The result for $dA_-$ is simply obtained by changing the sign of $\eta$ and replacing $\hat d^T\rightarrow\hat d$ in the above expression
\begin{equation}
dA_-=
\frac12\mathcal O_-^{-1}\{A_-,A_-\}
-\frac12c\eta^2\mathcal O_-^{-1}\{\hat dA_-,\hat dA_-\}
-\eta\mathcal O_-^{-1}R_g\{A_-,\hat dA_-\}\,.
\label{eq:dAm-eta}
\end{equation}
After rewriting $dA_+$ as in~\eqref{eq:dA-plus-eta} and projecting with $P^{(2)}$ we find
\begin{align}
dE^{(2)}=&\,
\{A_+^{(0)},E^{(2)}\}
+\frac12\{E^{(1)},E^{(1)}\}
+\frac12\{E^{(3)},E^{(3)}\}
-2\hat\eta\{E^{(3)},P^{(3)}\mathcal O_-^{-1}E^{(2)}\}
\nonumber\\
&{}
+4\hat\eta^{-1}P^{(2)}\mathcal O_+^{-1}\{E^{(2)},E^{(3)}\}
-8P^{(2)}\mathcal O_+^{-1}\{E^{(2)},P^{(3)}\mathcal O_-^{-1}E^{(2)}\}
\nonumber\\
&{}
+2\hat\eta^2\{P^{(3)}\mathcal O_-^{-1}E^{(2)},P^{(3)}\mathcal O_-^{-1}E^{(2)}\}
+2\eta\hat\eta^{-2}P^{(2)}\mathcal O_+^{-1}R_g\{E^{(2)},E^{(2)}\}\,,
\label{eq:dE2-eta}
\end{align}
where we have used (\ref{eq:supervielbeins}) to write the result in terms of the supervielbeins, together with (\ref{eq:Apm-rel}) and (\ref{eq:adh}).
We check again that the bosonic torsion $T^a$ takes the standard form~\eqref{eq:bos-torsion},
where we can now identify the spin connection for the $\eta$-model background as
\begin{equation}
\Omega_{ab}
=
(A_+)_{ab}
+2i\hat\eta(\gamma_{[a}E^2)_{\hat\alpha} M^{\hat\alpha2}{}_{b]}
+\frac{3i}{2}\hat\eta^2E^c\,M^{\hat\alpha2}{}_{[a}(\gamma_b)_{\hat\alpha\hat\beta}M^{\hat\beta2}{}_{c]}
-\frac12E^c(2M_{c[a,b]}-M_{ab,c})\,.
\label{eq:connection-eta}
\end{equation}
As before, we continue by computing the remaining components of the torsion.
First, from (\ref{eq:dA-minus-eta}) we get
\begin{align}
dE^{(3)}=&\,
\{A_+^{(0)},E^{(3)}\}
+\hat\eta P^{(3)}\mathcal O_-^{-1}\{E^{(3)},E^{(3)}\}
+2\{P^{(0)}\mathcal O_-^{-1}E^{(2)},E^{(3)}\}
\nonumber\\
&{}
+P^{(3)}(4\mathcal O_-^{-1}-1-2\hat\eta^{-2})\Ad_h^{-1}\{E^{(2)},E^{(1)}\}
-2\eta\hat\eta^{-1}P^{(3)}\mathcal O_-^{-1}R_g\Ad_h^{-1}\{E^{(2)},E^{(2)}\}
\nonumber\\
&{}
+2\hat\eta P^{(3)}(4\mathcal O_-^{-1}-1-2\hat\eta^{-2})\{\Ad_h^{-1}E^{(2)},P^1\mathcal O_-^{-1}E^{(2)}\}
\,,
%
\end{align}
which we use to check that also $T^{\hat\alpha2}$ is of the standard form~\eqref{eq:ferm2-torsion}.
To do this we make use of the spin connection (\ref{eq:connection-eta}) and we identify the following superfields for the $\eta$-model
\begin{align}
H_{abc}=&\,3M_{[ab,c]}-3i\hat\eta^2M^{\hat\alpha2}{}_{[a}(\gamma_b)_{\hat\alpha\hat\beta}M^{\hat\beta2}{}_{c]}\,,
\label{eq:Habc-eta}
\\
\mathcal S^{\hat\alpha1\hat\beta2}=&\,8i[\Ad_h(1+2\hat\eta^{-2}-4\mathcal O_+^{-1})]^{\hat\alpha1}{}_{\hat\gamma1}\widehat{\mathcal K}^{\hat\gamma1\hat\beta2}\,,
\label{eq:S-eta}
\\
\chi^2_{\hat\alpha}=&\,-\frac{i}{2}\hat\eta\gamma^a_{\hat\alpha\hat\beta}M^{\hat\beta2}{}_a\,,
\label{eq:chi2-eta}
\\
\psi_{ab}^{\hat\alpha2}=&
-2\eta\hat\eta^{-1}[\mathcal O_-^{-1}R_g\Ad_h^{-1}]^{\hat\alpha2}{}_{cd}\widehat{\mathcal K}_{ab}{}^{cd}
+\frac14\hat\eta[\Ad_hM]^{\hat\beta1}{}_{[a}(\gamma_{b]}\mathcal S^{12})_{\hat\beta}{}^{\hat\alpha}
\,.
\end{align}
To identify the last component of the spinor superfields we must compute torsion $T^{\hat\alpha1}$. Starting from (\ref{eq:dA-plus-eta}) we find
\begin{align}
dE^{(1)}
=&\,
\{\Ad_hA_+^{(0)}-dhh^{-1},E^{(1)}\}
+\hat\eta P^{(1)}\Ad_h\mathcal O_+^{-1}\Ad_h^{-1}\{E^{(1)},E^{(1)}\}
\nonumber\\
&{}
+P^{(1)}\Ad_h(4\mathcal O_+^{-1}-1-2\hat\eta^{-2})\{E^{(2)},E^{(3)}\}
+2\eta\hat\eta^{-1}P^{(1)}\Ad_h\mathcal O_+^{-1}R_g\{E^{(2)},E^{(2)}\}
\nonumber\\
&{}
-2\hat\eta P^{(1)}\Ad_h(4\mathcal O_+^{-1}-1-2\hat\eta^{-2})\{E^{(2)},P^{(3)}\mathcal O_-^{-1}E^{(2)}\}
\,.
\label{eq:dE1-eta}
\end{align}
Using this expression we can check\footnote{As in the previous section, we need to first find an expression for $\Ad_hA_+^{(0)}-dhh^{-1}$, see~\eqref{eq:dh-eta}.} that $T^{\hat\alpha1}$ is standard, see~\eqref{eq:ferm1-torsion}, where $\mathcal S^{\hat\beta2\hat\alpha1}=-\mathcal S^{\hat\alpha1\hat\beta2}$ and
\begin{align}
\chi^1_{\hat\alpha}=\frac{i}{2}\hat\eta\gamma^b_{\hat\alpha\hat\beta}[\Ad_hM]^{\hat\beta1}{}_b\,,
\label{eq:chi1-eta}
\qquad
\psi_{ab}^{\hat\alpha1}=
2\eta\hat\eta^{-1}[\Ad_h\mathcal O_+^{-1}R_g]^{\hat\alpha1}{}_{cd}\widehat{\mathcal K}_{ab}{}^{cd}
-\frac14\hat\eta(\mathcal S^{12}\gamma_{[a})^{\hat\alpha}{}_{\hat\beta}M^{\hat\beta2}{}_{b]}
\,.
\end{align}
Let us also note that in the case of the $\eta$-model the $B$-field can be written in the two ways
\begin{equation}\label{eq:B-eta}
B=\frac{\hat\eta^2}{4}\mathrm{Str}(g^{-1}dg\wedge\hat d^TA_+)
=-\frac{\hat\eta^2}{4}\mathrm{Str}(g^{-1}dg\wedge\hat dA_-)\,,
\end{equation}
which are equivalent thanks to the properties of $\mathcal O_\pm$ under transposition.

\subsection{Dilaton and supergravity condition}\label{sec:dilaton}
Unlike in the case of the $\lambda$-model, the $\eta$-model does not come with a natural candidate dilaton. Indeed, in general the target space geometry of the $\eta$-model is a solution of the generalized type II supergravity equations of~\cite{Arutyunov:2015mqj,Wulff:2016tju} rather than the standard ones, and a dilaton does not exist. One of our goals is to determine precisely when a dilaton exists for the $\eta$-model. To do this, let us define a would-be dilaton in the same way as the dilaton is defined in the $\lambda$-model
\begin{equation}\label{eq:e-2phi-eta}
e^{-2\phi}=\mathrm{sdet}(\mathcal O_+)=\mathrm{sdet}(1+\eta R_g\hat d^T)\,.
\end{equation}
For this to be the actual dilaton of the $\eta$-model its spinor derivatives must coincide with the dilatinos in (\ref{eq:chi2-eta}) and (\ref{eq:chi1-eta}). In~\eqref{eq:dphi-eta} we write down the result for $d\phi$. In particular we find\footnote{Here we used the fact that $\mathcal O_\pm^{-1}P^{(0)}=P^{(0)}$.}
\begin{align}
\nabla_{\hat\alpha2}\phi
=&
-2\hat\eta^{-1}\mathrm{STr}(P^a\mathcal O_+^{-1}[Q^2_{\hat\alpha},P_a])
-\frac{\eta}{2}\hat\eta^{-1}\widehat{\mathcal K}^{AB}\mathrm{STr}(T_AR_g[T_B,Q^2_{\hat\alpha}])
\nonumber\\
=&
\chi^2_{\hat\alpha}
-\frac{\eta}{2}\hat\eta^{-1}\widehat{\mathcal K}^{AB}\mathrm{STr}([T_A,RT_B]gQ^2_{\hat\alpha} g^{-1})\,,
\label{eq:dalpha2phi}
\\
\nabla_{\hat\alpha1}\phi
=&
-\hat\eta[\Ad_h^{-1}]^{\hat\beta}{}_{\hat\alpha}
(
\widehat{\mathcal K}^{\hat\gamma1\hat\delta2}\mathrm{STr}(Q^2_{\hat\delta}\mathcal O_+^{-1}[Q^1_{\hat\beta},Q^1_{\hat\gamma}])
-\frac{\eta}{2}\widehat{\mathcal K}^{AB}\mathrm{STr}(T_AR_g[T_B,Q^1_{\hat\beta}])
)
\nonumber\\
=&
\chi^1_{\hat\alpha}
+\frac{\eta}{2}\hat\eta[\Ad_h^{-1}]^{\hat\beta}{}_{\hat\alpha}\widehat{\mathcal K}^{AB}\mathrm{STr}([T_A,RT_B]gQ^1_{\hat\beta} g^{-1})\,.
\label{eq:dalpha1phi}
\end{align}
Therefore a sufficient condition for the $\eta$-model to lead to a standard supergravity background is that
\begin{equation}
\widehat{\mathcal K}^{AB}\mathrm{STr}([T_A,RT_B]gQ^I_{\hat\alpha} g^{-1})=0
%
%
\,,
\end{equation}
or, since $g$ is an arbitrary group element (modulo gauge-transformations),
\begin{equation}
\mathrm{STr}(R\mathrm{ad}_x)=0\,,\qquad\forall x\in\mathfrak{g}\qquad(\mbox{i.e.}\ \ R^B{}_Af^A{}_{BC}=0, \text{ or }R^{BC}f^A{}_{BC}=0)\,.
\label{eq:sugra-cond}
\end{equation}
To see that this condition is also necessary we calculate the Killing vector superfield $K^a$ appearing in the generalized supergravity equations of~\cite{Wulff:2016tju}, which in general is given by
\begin{equation}\label{eq:Ka-dchi}
K^a=-\frac{i}{16}(\gamma^a)^{\hat\alpha\hat\beta}(\nabla_{\hat\alpha1}\chi_{\hat\beta1}-\nabla_{\hat\alpha2}\chi_{\hat\beta2})\,,
\end{equation}
and whose result is collected in~\eqref{eq:Ka}.
The $\eta$-model has a standard type II supergravity solution as target space if $K^a=0$. In fact, it must be that it vanishes order by order in the deformation parameter $\eta$. At linear order we find the equation
\begin{equation}
\widehat{\mathcal K}^{AB}\mathrm{STr}([T_A,RT_B]gP_ag^{-1})=0\,,
\end{equation}
which, since $g\in G$ is arbitrary implies (\ref{eq:sugra-cond}). Therefore the condition (\ref{eq:sugra-cond}) is both necessary and sufficient, and also the higher order terms in $\eta$ in (\ref{eq:Ka}) vanish when this condition is fulfilled.

\section{Non-abelian $R$-matrices and the unimodularity condition}\label{sec:R-matrices}
In this section we study the unimodularity condition (\ref{eq:uni-R}) for the $R$-matrix.
First we analyse its compatibility with a class of non-abelian $R$-matrices---the Jordanian ones---and then we explain how to classify all unimodular $R$-matrices solving the CYBE on the bosonic subalgebra of the superisometry algebra.

Following~\cite{2004math......2433T} we define an ``extended Jordanian'' $R$-matrix for a Lie superalgebra $\mathfrak g$ as follows: we fix a Cartan element $h$ ($\deg(h)=0$) and a positive root $e$ as well as a collection of roots $e_{\gamma_{\pm i}}$ with $i\in\{1,2,\ldots,N\}$ such that $\deg(e)=\deg(e_{\gamma_i})+\deg(e_{\gamma_{-i}})$ (mod 2) and satisfying
\begin{align}
&[h,e]=e\,,\qquad
[h,e_{\gamma_i}]=(1-t_{\gamma_i})e_{\gamma_i}\,,\qquad
[h,e_{\gamma_{-i}}]=t_{\gamma_i}e_{\gamma_{-i}}\,,\qquad (t_{\gamma_i}\in\mathbbm{C})
\nonumber\\
&[e_{\gamma_{\pm i}},e]=0\,,\qquad[e_{\gamma_k},e_{\gamma_l}]=\delta_{k,-l}e\,,\qquad (k>l\in\{\pm1,\pm2,\ldots,\pm N\})\,.
\label{eq:cond-ext-jord}
\end{align}
The extended Jordanian $R$-matrix is then defined as
\begin{equation}
R=h\wedge e+\sum_{i=1}^N(-1)^{\deg(e_{\gamma_i})\deg(e_{\gamma_{-i}})}e_{\gamma_i}\wedge e_{\gamma_{-i}}\,.
\end{equation}
It is now easy to see that for a bosonic deformation, i.e. $\deg(e)=0$, we have
\begin{equation}
r^{ij}[b_i,b_j]=(N_0-N_1+1)e\,,
\end{equation}
with $N=N_0+N_1$, $N_0$ ($N_1$) being the number of bosonic (fermionic) roots $e_{\gamma_i}$. For this to vanish we need precisely one more fermionic $e_{\gamma_i}$ than bosonic. This is clearly a very strong restriction on the allowed Jordanian $R$-matrices. Let us note that this result is compatible with the findings of~\cite{Kyono:2016jqy,Hoare:2016hwh,Orlando:2016qqu}, where Jordanian $R$-matrices acting only on bosonic generators were found to produce backgrounds which do not solve the standard supergravity equations. We have considered certain examples of bosonic Jordanian $R$-matrices (namely $R=J_{01}\wedge (P_0-P_1)$, $R=J_{03}\wedge (J_{01}-J_{13})$ and $R=D\wedge p_i,\ i=0,\ldots,3$) and we have checked that it is not possible to find a positive and a negative fermionic root satisfying~\eqref{eq:cond-ext-jord} without spoiling the reality of the extended $R$-matrix. If possible, it would be interesting to find extended Jordanian unimodular $R$-matrices for $\mathfrak{psu}(2,2|4)$, but we will not analyze this question further here.

From now on we will restrict to the bosonic subalgebra $\mathfrak{so}(2,4)\oplus\mathfrak{so}(6)\subset\mathfrak{psu}(2,2|4)$. Let us recall some known facts about solutions to the CYBE, (\ref{eq:YBE}) with $c=0$, for ordinary Lie algebras. The first important fact, due to Stolin~\cite{Stolin1991a,STOLIN1999285}, is that there is a one-to-one correspondence between constant solutions of the CYBE for a Lie algebra $\mathfrak g$ and quasi-Frobenius (or symplectic) subalgebras $\mathfrak f\subset\mathfrak g$ (see also~\cite{Gerstenhaber1997}). Notice that we do not need to assume anything about the Lie algebra $\mathfrak g$, in particular it does not need to be simple. A Lie algebra is quasi-Frobenius if it has a non-degenerate 2-cocycle $\omega$, i.e.
\begin{equation}\label{eq:2cocycle}
\omega(x,y)=-\omega(y,x)\,,\qquad\omega([x,y],z)+\omega([z,x],y)+\omega([y,z],x)=0\,,\qquad\forall x,y,z\in\mathfrak f \,.
\end{equation}
It is Frobenius if $\omega$ is a coboundary, i.e. $\omega(x,y)=f([x,y])$ for some linear function $f$. If $R$ is a solution to the CYBE for $\mathfrak g$, then there is a  subalgebra $\mathfrak f$ on which $R$ is non-degenerate. This subalgebra is necessarily quasi-Frobenius, and writing $R$ in the form (\ref{eq:R-exp}) the 2-cocycle is the inverse of the $R$-matrix, i.e. $\omega(b_i,b_j)=(r^{-1})_{ij}$. The converse is also true, i.e. if $\mathfrak f\subset\mathfrak g$ is quasi-Frobenius then the inverse of the 2-cocycle $\omega$ gives a solution to the CYBE, as is easily verified. Therefore,  finding solutions to the CYBE for a given $\mathfrak g$ reduces to finding all quasi-Frobenius subalgebras\footnote{This was done for $\mathfrak{sl}(2)$ and $\mathfrak{sl}(3)$ in~\cite{Stolin1991b}. } of $\mathfrak g$. 
A fact with important consequences for our analysis is that if $\mathfrak g$ is compact then $\mathfrak f$ must be abelian~\cite{Lichnerowicz1988}.
This leads to the conclusion that deformations involving only $S^5$ (i.e. marginal deformations of the dual CFT) must necessarily have abelian $R$-matrices.

We now show that the unimodularity condition (\ref{eq:uni-R}) for the $R$-matrix adds a further property to the quasi-Frobenius subalgebra $\mathfrak f$. If we write the structure constants as $f^i{}_{jk}$ in some basis, the 2-cocycle condition is
\begin{equation}
(r^{-1})_{i[j}f^i{}_{kl]}=0\,.
\end{equation}
Contracting this equation with $r^{jk}$ we get $(r^{-1})_{il}f^i{}_{jk}r^{jk}=-2f^i{}_{il}$, which
together with the unimodularity condition  for the $R$-matrix written as (\ref{eq:bb-cond}), i.e. $f^i{}_{jk}r^{jk}=0$,  implies
\begin{equation}
f^i{}_{il}=0\quad\Leftrightarrow\quad\mathrm{tr}(ad_x)=0\quad\forall x\in\mathfrak f\,.
\end{equation}
Therefore  $\mathfrak f$ is a \emph{unimodular} Lie algebra.  Clearly the converse is also true and we have established that

\noindent
\textit{
Solutions of the CYBE for a Lie algebra $\mathfrak g$ which satisfy the condition (\ref{eq:uni-R}) are in one-to-one correspondence with unimodular quasi-Frobenius subalgebras of $\mathfrak g$.
}

For this reason we refer also to the $R$-matrices which satisfy (\ref{eq:uni-R}) as unimodular.
A quasi-Frobenius Lie algebra must clearly have even dimension, and if the dimension is two the algebra must be abelian to respect unimodularity. To find a non-abelian $R$-matrix we must therefore consider at least the case of rank four. Luckily the real quasi-Frobenius Lie algebras of dimension four were classified in~\cite{Ovando2006}, and the five unimodular ones (Corollary 2.5 in~\cite{Ovando2006}) are listed in table~\ref{tab:algebras}.
\begin{table}[ht]
\begin{center}
\begin{tabular}{c|c}
$\mathfrak{f}$ & Defining Lie brackets \\
\hline
$\mathbbm{R}^4$ & -- \\
$\mathfrak{h}_3\oplus\mathbbm{R}$ & $[e_1,e_2]=e_3$ \\
$\mathfrak{r}_{3,-1}\oplus\mathbbm{R}$ & $[e_1,e_2]=e_2$, $[e_1,e_3]=-e_3$ \\
$\mathfrak{r}'_{3,0}\oplus\mathbbm{R}$ & $[e_1,e_2]=-e_3$, $[e_1,e_3]=e_2$ \\
$\mathfrak{n}_4$ & $[e_1,e_2]=-e_4$, $[e_4,e_2]=e_3$ 
\end{tabular}
\end{center}
\caption{The four-dimensional real unimodular quasi-Frobenius Lie algebras. In all cases the 2-cocycle can be taken as  $\omega=e^1\wedge e^4+e^2\wedge e^3$, where $e^i$ denotes the dual basis of $\mathfrak f^*$.}
\label{tab:algebras}
\end{table}
The task of finding all $R$-matrices of rank four which solve the CYBE and lead to a deformation of the $AdS_5\times S^5$ string with a proper supergravity background is therefore reduced to finding all inequivalent embeddings of these subalgebras in $\mathfrak{so}(2,4)\oplus\mathfrak{so}(6)$. The most interesting problem is to find the embedding of the non-abelian algebras\footnote{The extension to $\mathfrak{so}(2,4)\oplus\mathfrak{so}(6)$ is essentially trivial and amounts to adding in commuting generators from $\mathfrak{so}(6)$ in such a way that the commutation relations of the algebra are preserved.} in $\mathfrak{so}(2,4)$.  This is still quite challenging, but it becomes simpler by the following observation. 
A unimodular quasi-Frobenius Lie algebra is solvable~\cite{Lichnerowicz1988}, and solvable subalgebras of $\mathfrak{so}(2,4)$ must be embeddable in one of the maximal solvable subalgebras of $\mathfrak{so}(2,4)$, see~\cite{Patera:1973yn} for a proof of this.  Besides the Cartan subalgebra which is not relevant for our purposes, Patera, Winternitz and Zassenhaus in~\cite{Patera1973b} showed that there are two maximal solvable subalgebras of $\mathfrak{so}(2,4)$, $\mathfrak s_1$ and $\mathfrak s_2$ of dimension 9 and 8 respectively. It is most convenient to write them using the conformal form of the $\mathfrak{so}(2,4)$ algebra, with dilatation generator $D$, translations and special conformal generators $p_i,\, k_i$ ($i=0,\ldots 3$) and Lorentz transformations and rotations $J_{ij}$. They are related to the form of $\mathfrak{so}(2,4)$ in (\ref{eq:alg-b}) with $\widehat{\mathcal K}_{ij}{}^{kl}=-2\delta_{[i}^k\delta_{j]}^l$ by
\begin{equation}\label{eq:def-cnf-gen}
p_i=P_i+J_{i4}\,,\qquad k_i=-P_i+J_{i4}\,,\qquad D=P_4\,,
\end{equation}
and the non-vanishing commutators are
\begin{align}
&[D,p_i]=p_i\,,\qquad [D,k_i]=-k_i\,,\qquad [p_i,k_j]=-2\eta_{ij}D+2J_{ij}\,,
\\
&[J_{ij},p_k]=2\eta_{k[i}p_{j]}\,,\qquad[J_{ij},k_k]=2\eta_{k[i}k_{j]}\,,\qquad
[J_{ij},J_{kl}]=\eta_{ik}J_{jl}-\eta_{jk}J_{il}-\eta_{il}J_{jk}+\eta_{jl}J_{ik}\,.
\nonumber
\end{align}
The metric on the Lie algebra is given by $\mathrm{tr}(DD)=1$, $\mathrm{tr}(p_ik_j)=-2\eta_{ij}$, $\mathrm{tr}(J_{ij}J_{kl})=-2\eta_{i[k}\eta_{l]j}$\,. The two non-abelian maximal solvable subalgebras of $\mathfrak{so}(2,4)$ then take the form
\begin{align}
&\mathfrak{s}_1=\mathrm{span}(p_i,\,J_{01}-J_{13},\,J_{02}-J_{23},\,J_{03},\,J_{12},\,D)\,,
\nonumber\\
&\mathfrak{s}_2=\mathrm{span}(p_0+p_3,\,p_1,\,p_2,\,J_{01}-J_{13},\,J_{02}-J_{23},\,J_{12},\,J_{03}-D,\,k_0+k_3+2p_3)\,,
%
%
%
%
%
%
\end{align}
up to automorphisms. Our task is reduced to finding all embeddings of the non-abelian algebras in table \ref{tab:algebras} in $\mathfrak{s}_1$ and $\mathfrak{s}_2$. To simplify this problem further we will single out the element $e_3$ in this table\footnote{The reason for picking $e_3$  is that it always arises as a commutator of two other elements. Since the last three generators in $\mathfrak{s}_1$ or $\mathfrak{s}_2$ are never generated in commutators, they do not appear in $e_3$.} and use automorphisms generated by elements of $\mathfrak{s}_1$ ($\mathfrak{s}_2$) to simplify it as much as possible.  Using this freedom we can bring $e_3$ to one of the following forms
\begin{align}
\mathfrak{s}_1:\qquad&
(1)\quad e_3=p_1\,,\qquad
(2)\quad e_3=J_{02}-J_{23}\,,\qquad
(3)\quad e_3=p_1+J_{02}-J_{23}\,,\qquad
(4)\quad e_3=p_0\,,
\nonumber\\
&(5)\quad e_3=p_3\,,\qquad
(6)\quad e_3=p_0+p_3\,,\qquad
(7)\quad e_3=p_0-p_3+J_{01}-J_{13}\,,
\\
\mathfrak{s}_2:\qquad&
(1)\quad e_3=p_1\,,\qquad
(2)\quad e_3=p_0+p_3\,,\qquad
(3)\quad e_3=ap_1+bp_2+J_{01}-J_{13}\,.
%
\end{align}
The rest is a straightforward if slightly tedious calculation. The results are summarized in tables \ref{tab:h3}--\ref{tab:n4}. Note that in writing these embeddings we have used automorphisms of the four-dimensional subalgebras which are not always inner automorphisms of $\mathfrak{so}(2,4)$. This must be accounted for when constructing the list of inequivalent $R$-matrices.
In Table \ref{tab:R-matrices} in the introduction we write the corresponding $R$-matrices, $R=e_1\wedge e_4+e_2\wedge e_3$ up to 
automorphisms. 
In Table \ref{tab:R-matrices-inner} instead we list the inequivalent, modulo inner automorphisms of $\mathfrak{so}(2,4)$, $R$-matrices. This is the result which is interesting from the string sigma model perspective, since inner automorphisms correspond to field redefinitions in the sigma model, i.e. coordinate transformations in target space.
In Table \ref{tab:R-matrices-iso-inner} we write down the bosonic isometries and the number of supercharges that each $R$-matrix preserves.  Given a generator $t$ of the superalgebra $\mathfrak g$, the condition that it is preserved by the $R$-matrix is given by
\begin{equation}
[t,R(x)]=R([t,x])\,,\qquad\forall x\in\mathfrak g\,.
\label{eq:isometries}
\end{equation}

Most of these $R$-matrices all have a form which suggests that they should correspond to non-commuting TsT-transformations\footnote{Here we use TsT in a generalized sense, where we can involve also non-compact directions}, in the sense that they involve sequences of T-dualities along non-commuting directions. All but the last three $R$-matrices in table \ref{tab:R-matrices} have the form
\begin{equation}
R=a\wedge b+c\wedge d\,,
\end{equation}
where $[a,b]=[c,d]=0$ and $c,d$ generate isometries of the corresponding background. It is natural to conjecture that such $R$-matrices correspond to two successive TsT-transformations, the first using isometries $a,b$ and the second using isometries $c,d$. Note that unlike in standard applications of TsT-tranformations, e.g.~\cite{Alday:2005ww}, the pairs of isometries $a,b$ and $c,d$ do not commute with each other. 
This means that after the first TsT is implemented, it is necessary to make a change of coordinates in order to realize the isometries of the second TsT transformation as shift isometries. We will confirm this in section \ref{sec:backgrounds}, when we will check in some examples that the deformed backgrounds are indeed equivalent to such sequences of TsT-transformations. These considerations suggest a very simple picture for how TsT-transformations are interpreted at the level of the $R$-matrix: the TsT-transformation involving isometries $a,b$ should be simply implemented by adding a term $a\wedge b$ to the $R$-matrix\footnote{It is easy to check that this is compatible with the CYBE, since $a,b$ are isometries and satisfy~\eqref{eq:isometries}.}. Notice that the number of free parameters entering the definitions of the $R$-matrices (plus the overall deformation parameter) does not need to be equal to the number of TsT-transformations implemented. In fact, the number of parameters could be reduced in some cases, if they can be reabsorbed by means of field redefinitions. In other cases one might have more parameters than expected, which suggests the possibility of applying TsT-transformations on linear combinations of the isometric coordinates.

The structure of the last three $R$-matrices in table \ref{tab:R-matrices} is different, and one observes that now $a,c$ generate isometries. However, one can check explicitly that the background corresponding to $R_{15}$, for example, is self-dual (up to field redefinitions) under a TsT-transformation involving $a,c$.\footnote{Note that this is consistent with our above proposal on how to interpret the action of TsT at the level of the $R$-matrix; in fact, in this case the addition of the term $a\wedge c$ to $R_{15}$ can be removed by an inner automorphism of $\mathfrak{so}(2,4)$. Here $a,c$ can be chosen to be $p_1,p_0+p_3$.} This example is particularly instructive because it can be embedded in $\mathfrak{so}(2,3)$: in this algebra, the deformed background does not preserve other bosonic isometries than $a,c$, which suggests that backgrounds corresponding to the algebra $\mathfrak n_4$ are not of TsT-type. Note that $\mathfrak n_4$ is the only algebra considered which is not the direct sum of a three-dimensional algebra and a commuting generator. One possibility is that \emph{non-abelian} T-duality of the corresponding subalgebra should instead play a role in the interpretation of these backgrounds. A hint towards this direction comes from the results of~\cite{Elitzur:1994ri}, where it was shown that a conformal anomaly is encountered when implementing non-abelian T-duality on a subalgebra, unless all generators have vanishing trace.\footnote{We thank Arkady Tseytlin for pointing this reference out to us.} In the case of the adjoint representation this condition is precisely that of unimodularity of the corresponding subalgebra.

\begin{table}[ht]
$$
\begin{array}{c|cccc}
\mathfrak h_3\oplus\mathbbm{R} & e_1 & e_2 & e_3 & e_4\\
\hline
1. & p_1 & J_{01}-J_{13} & p_0+p_3 & p_2\\
2. & p_1 & p_3+J_{01}-J_{13} & p_0+p_3 & p_2\\
3. & p_1 & p_2+J_{01}-J_{13} & p_0+p_3 & p_1+J_{02}-J_{23}\\
4. & \frac12p_1-\frac12(J_{02}-J_{23}) & p_2+J_{01}-J_{13} & p_0+p_3 & k_0+k_3+2p_3-2J_{12}
\end{array}
$$
\caption{Embeddings of $\mathfrak h_3\oplus\mathbbm{R}$ in $\mathfrak{so}(2,4)$ up to automorphism.}
\label{tab:h3}
\end{table}

\begin{table}[ht]
$$
\begin{array}{c|cccc}
\mathfrak r_{3,-1}\oplus\mathbbm{R} & e_1 & e_2 & e_3 & e_4\\
\hline
1. & -D-J_{03} & J_{02}-J_{23} & p_1 & p_0+p_3\\
2. & J_{03} & p_0-p_3 & p_0+p_3 & p_1\\
3. & J_{03} & p_0-p_3 & p_0+p_3 & J_{12}\\
(4.) & D+2J_{03} & p_1 & p_0+p_3 & -
\end{array}
$$
\caption{Embeddings of $\mathfrak r_{3,-1}\oplus\mathbbm{R}$ in $\mathfrak{so}(2,4)$ up to automorphism. The last case is an embedding of $\mathfrak r_{3,-1}$ which does not extend to an embedding of $\mathfrak r_{3,-1}\oplus\mathbbm{R}$. It is the only case where this happens and included only since it is relevant for constructing all non-abelian $R$-matrices of $\mathfrak{so}(2,4)\oplus\mathfrak{so}(6)$.}
\label{tab:r3-1}
\end{table}

\begin{table}[ht]
$$
\begin{array}{c|cccc}
\mathfrak r'_{3,0}\oplus\mathbbm{R} & e_1 & e_2 & e_3 & e_4\\
\hline
1. & J_{12} & p_2 & p_1 & p_0+p_3\\
2. & p_3+J_{12} & p_2 & p_1 & p_0+p_3\\
3. & p_0+J_{12} & p_2 & p_1 & p_3\\
4. & J_{12} & p_2 & p_1 & p_3\\
5. & p_3+J_{12} & p_2 & p_1 & p_0\\
6. & J_{12} & p_2 & p_1 & p_0\\
7. & J_{12} & p_2 & p_1 & J_{03}
\end{array}
$$
\caption{Embeddings of $\mathfrak r'_{3,0}\oplus\mathbbm{R}$ in $\mathfrak{so}(2,4)$ up to automorphism.}
\label{tab:r30}
\end{table}

\begin{table}[ht]
$$
\begin{array}{c|cccc}
\mathfrak n_4 & e_1 & e_2 & e_3 & e_4\\
\hline
1. & p_3 & J_{01}-J_{13} & p_0+p_3 & p_1\\
2. & p_3 & p_2+J_{01}-J_{13} & p_0+p_3 & p_1\\
3. & p_1+p_3+J_{02}-J_{23} & p_2+J_{01}-J_{13} & p_0+p_3 & p_1
\end{array}
$$
\caption{Embeddings of $\mathfrak n_4$ in $\mathfrak{so}(2,4)$ up to automorphism.}
\label{tab:n4}
\end{table}

Let us now consider the case of higher ranks, which can only be six or eight.
We have not done a systematic study for the case of rank six $R$-matrices. One would first need to identify all 6-dimensional subalgebras of $\mathfrak s_1$ and $\mathfrak s_2$, and check which of them are unimodular and quasi-Frobenius. We have found that the subalgebra of $\mathfrak s_1$ generated by $\{ p_i, J_{03}, J_{12} \}$ has both properties. It is straightforward to find the 2-form $\omega$ that solves the cocycle condition~\eqref{eq:2cocycle}, and invert it to find the corresponding $R$-matrix. For particular choices of the free parameters this can be written e.g. as $R=p_0\wedge p_1 + p_2\wedge p_3 + J_{01}\wedge J_{23}$.

We have also checked that there is no 8-dimensional subalgebra which is at the same time unimodular and quasi-Frobenius. Therefore there is no rank eight $R$-matrix which produces a background that solves the supergravity equations of motion.
It is in fact easy to check that $\mathfrak s_2$ (which is 8-dimensional) is quasi-Frobenius but not unimodular.
To identify all 8-dimensional subalgebras of $\mathfrak s_1$ (which is 9-dimensional), we first define $e=\sum_{j=1}^9 \lambda_j e_j$ to be the generator which we want to remove, where $e_j$ are the generators of $\mathfrak s_1$.
Then for a generic element $X\in\mathfrak s_1$ we define its component perpendicular to $e$ as $X^\perp = X-P(X)$, where $P$ projects\footnote{We define $P(X) = e \, \text{STr}(X e^*)$, where $e^*$ is a dual to $e$, $\text{STr}(ee^*)=1$. We can take it as $e^* = \sum_{j=1}^9 \tfrac{\lambda_j}{||\lambda||^2} e^j$, where $||\lambda||^2=\sum_{j=1}^9 \lambda_j^2$ and $e^j$ are the duals of the generators in the basis such that $\text{STr}(e_ie^j)=\delta_i^j$.} along $e$.
Then the condition to have a subalgebra is $P[X^\perp,Y^\perp ]=0, \ \forall X,Y\in \mathfrak s_1$.
These equations give two possible solutions, depending on some unconstrained parameters
\begin{equation}
\begin{aligned}
(a) \qquad & e= \lambda_7 J_{12} + \lambda_8 J_{03}+\lambda_9 D\,,\\
(b)\qquad  & e= \lambda_1 (p_0-p_3) + \lambda_8 (J_{03}- D)\,.\\
\end{aligned}
\end{equation}
In the case $(a)$ we find\footnote{In both cases $(a)$ and $(b)$ one needs to choose carefully a basis for the 8-dimensional subalgebra, in such a way that the generators are linearly independent and non-degenerate for generic choices of the remaining $\lambda_j$. A way to do it is to pick an orthogonal basis, and normalise the vectors such that they can be degenerate only if $\lambda_j=0\ \forall j$.} that the subalgebra is unimodular if $\lambda_7=0$ and $\lambda_9=2\lambda_8$. However, for this choice it is not quasi-Frobenius---the cocycle condition gives a 2-form of rank six.
In the case $(b)$ the subalgebra is not unimodular for any choice of $\lambda_1,\ \lambda_8$.

\section{Some examples of supergravity backgrounds}\label{sec:backgrounds}
In this section we give a brief discussion on the $\eta$-model backgrounds generated by solutions of the CYBE ($c=0$), when we restrict $R$ to act only on the bosonic subalgebra.
In most cases a convenient parameterisation of the group element $g=g_a\cdot g_s \in SO(2,4)\times SO(6)$ is
\begin{equation}
g_a = \exp \left( 
x^i p_i \right) \cdot \exp \left( \log z \, D \right)\,,
\end{equation}
where $p_i,D$ are the generators defined in~\eqref{eq:def-cnf-gen}.
Here we will be interested only on deformations of AdS, so we will not need to specify the parameterisation that we use for $g_s$ on the sphere.
In this coordinate system the undeformed metric takes the familiar form
\begin{equation}
ds^2_{\eta=0} = \frac{\eta_{ij}dx^i\, dx^j+dz^2}{z^2}+ds_s^2\,.
\end{equation}
Because of our restriction on $R$, it is enough to look at the action of the operators $\mathcal O_\pm$ on the bosonic subalgebra.
They take a block form
\begin{equation}
\left(
\begin{array}{c|c}
\ \mathbf{1}\  & {(\mathcal O_\pm)^{bc}}_a \\
\hline
\ \mathbf{0}\  & {(\mathcal O_\pm)^{b}}_a
\end{array}
\right)\,,
\end{equation}
since\footnote{We recall that $P^{(0)}$ and $P^{(2)}$ are projectors on the subspaces spanned by the generators $J_{ab}$ and $P_a$ respectively. A useful matrix realisation of the algebra generators can be found in~\cite{Arutyunov:2015qva}. Here we identify $P_a=\mathbf{P}_a,$ and $J_{ab}=-\mathbf{J}_{ab}$, where $\mathbf{P}_a,\ \mathbf{J}_{ab}$ are the generators used in~\cite{Arutyunov:2015qva}.} $\mathcal O_\pm P^{(0)}= P^{(0)}$.
All the information about background fields of the deformed model can be extracted by studying just the block ${(\mathcal{O}_+)^b}_a$---or in other words $P^{(2)}\mathcal{O}_+ P^{(2)}$.
Notice that the results for ${(\mathcal{O}_-)^b}_a$ are simply obtained by changing the sign of the deformation parameter $\eta$.
The dilaton of the deformed model is easily obtained by computing the determinant of ${(\mathcal{O}_+)^b}_a$
\begin{equation}
e^\phi = (\det \mathcal O_+)^{-1/2}\,.
\end{equation}
The rest of the background fields are written in terms of ${(\mathcal{O}_+^{-1})^b}_a$---the inverse of the block ${(\mathcal{O}_+)^b}_a$.
The vielbein components for the deformed model are
\begin{equation}
E^a = {(\mathcal{O}_+^{-1})^a}_b e^b\,,
\end{equation}
where $e^a$ is the bosonic vielbein of the \emph{undeformed} background, related to the Maurer-Cartan form as
\begin{equation}
g^{-1}dg = e^a P_a + \tfrac{1}{2} \omega^{ab}J_{ab}\,.
\end{equation}
The spacetime metric of the deformed background is then straightforwardly obtained, $ds^2 = \eta_{ab}E^a E^b$.
The $B$-field can be extracted immediately from the action of the bosonic $\sigma$-model, and it reads as
\begin{equation}
B= \tfrac{1}{2} dX^n \wedge dX^m B_{mn}=\tfrac{1}{2} (\mathcal O_-^{-1})_{ab} \ e^a \wedge e^b\,,
\end{equation}
where it is assumed that indices are raised and lowered with $\eta_{ab}$.
To get the Ramond-Ramond fields we first need to consider the local Lorentz transformation given by $M$ in~\eqref{eq:def-M} and write its action on spinors
\begin{equation}
{(\Ad_h)^{\hat\beta}}_{\hat\alpha} = \exp[- \tfrac{1}{4} (\log M)_{ab} \Gamma^{ab} ]{^{\hat\beta}}_{\hat\alpha}\,,
\end{equation}
where here we have introduced a basis for $32\times 32$ Gamma-matrices\footnote{For a convenient basis see~\cite{Arutyunov:2015qva}.}.
The RR fields are obtained by solving the equation (note that (\ref{eq:RR-intro}) simplifies considerably for $R$-matrices of the bosonic subalgebra)
\begin{equation}
(\Gamma^a F_a + \tfrac{1}{3!} \Gamma^{abc} F_{abc} + \tfrac{1}{2\cdot 5!} \Gamma^{abcde} F_{abcde})\Pi = 
e^{-\phi} \ \Ad_h(-4 \Gamma_{01234})\Pi
\end{equation}
where $\Pi = \tfrac{1}{2} (1-\Gamma_{11})$ is a projector and $(-4 \Gamma_{01234})\Pi$ encodes the 5-form flux of the undeformed model.
The various components of $F$'s are found by multiplying the above equation by the relevant Gamma-matrix $\Gamma_{a_1\ldots a_{2m+1}}$ and then taking the trace.
This computation\footnote{For $F^{(5)}$ it is enough to look at half of the components, \emph{e.g.} $F_{0bcde}$, and construct the corresponding form $f^{(5)}$. Then $F^{(5)}=(1+*)f^{(5)}$, such that $F^{(5)}=*F^{(5)}$.} yields the $F$'s expressed with tangent indices, which are translated into form language by
$F^{(2m+1)}=\frac{1}{(2m+1)!}E^{a_{2m+1}}\wedge\ldots\wedge E^{a_1} F_{a_1\ldots a_{2m+1}}$.

In the rest of this section we present some backgrounds solving the standard supergravity equations which we have derived by using the above procedure.
We work out one example for each of the 4-dimensional non-abelian subalgebras  in table~\ref{tab:algebras}.

In section~\ref{sec:R-matrices} we have argued that the $R$-matrices related to the subalgebras $\mathfrak{h}_3 \oplus \mathbbm{R}$, $\mathfrak r'_{3,0}\oplus \mathbbm{R}$, $\mathfrak r_{3,-1}\oplus \mathbbm{R}$ should produce backgrounds which can be obtained by sequences of TsT-transformations starting from $AdS_5\times S^5$. We check this explicitly for the backgrounds that we have derived, where we follow the conventions of~\cite{Hoare:2015wia} for the T-duality rules~\cite{Bergshoeff:1995as,Green:1996bh,Hassan:1999bv}.
Because the isometries of the first TsT do not commute with those of the second one, we will see that before doing the last step it is necessary to implement a coordinate transformation, which realizes the second pair of isometries as shifts of the corresponding coordinates.
Let us mention that since we have chosen to have just one overall deformation parameter $\eta$ (i.e. we fix some free parameters in the definitions of the possible $R$-matrices), the shifts of the two TsT-transformations are related to each other. This does not need to be true for generic cases.

\subsection{$\mathfrak{h}_3 \oplus \mathbbm{R}$}
Let us choose the $R$-matrix (this corresponds to $R_1$ in table~\ref{tab:R-matrices} with $x^1\leftrightarrow x^3$)
\begin{equation}
R = (J_{03}+J_{13})\wedge (p_0+p_1)+p_2\wedge p_3\,,
\end{equation}
which preserves 4 bosonic isometries
\begin{equation}
p_2\,, \quad p_3\,, \quad p_0+p_1\,,\quad p_0-p_1-2(J_{02}+J_{12})\,,
\end{equation}
and 8 supercharges.
Clearly, it is convenient to introduce lightcone coordinates $x^\pm = x^0 \pm x^1$, since a shift of $x^+$ will correspond to an isometry.
The spacetime metric that we obtain is
\begin{equation}
\begin{aligned}
ds^2&=z^{-2}\left(1+\frac{4 \eta ^2}{z^4}\right)^{-1} \left(4 \eta ^2 z^{-4} { x^-} d { x^-} (2 d { x_2}- { x^-}
   d { x^-})+d { x_2}^2+d { x_3}^2\right)\\
&+\frac{-d { x^-} d { x^+}+dz^2}{z^2}+ds_s^2.
\end{aligned}
\end{equation}
The dilaton depends only on the $z$-coordinate, while the $B$-field also on $x^-$
\begin{equation}
e^\phi = \left(1+\frac{4 \eta ^2}{z^4}\right)^{-1/2},
\qquad\qquad
B=\frac{2 \eta   ( d { x_2}-{ x^-} d { x^-}) \wedge d { x_3}}{ \left(4 \eta ^2+z^4\right)}.
\end{equation}
The RR-fluxes turn out to be quite simple
\begin{equation}
F^{(5)}=(1+*)\frac{2 d {x^-}\wedge d  {x^+}\wedge d  {x_2}\wedge d  {x_3}\wedge d z}{z(z^4+4 \eta ^2)},
\qquad
F^{(3)}=\frac{4 \eta}{z^5}   (2  {x^-} d {x_2}-d {x^+})\wedge d {x^-}\wedge  dz.
\end{equation}
In order to show that this background can be obtained by a sequence of TsT-transformations, we start from the deformed background and show that we can reach the undeformed $AdS_5 \times S^5$ by TsT-transformations.
We will write $T(x_i)$ to indicate that we apply T-duality along the isometric coordinate $x_i$, and denote by $\tilde x_i$ the dual coordinate.
In this case we need to do the sequence
\begin{equation}
T(x_2),\ x_3\to x_3-2\eta \tilde x_2,\ T(\tilde x_2),
\qquad\qquad
T(\psi),\ w^+\to w^+-2\eta \tilde\psi,\ T(\tilde\psi),
\end{equation}
where we need to redefine the coordinates in the $013$ space
\begin{equation}
x^+ = 2(\psi^2 w^- + w^+),\quad
x^- = 2w^-,\quad
x_3 = -2\psi w^-,
\end{equation}
before applying the last TsT-transformation.
Obviously, starting from $AdS_5\times S^5$ and applying these TsT-transformations backwards, we find the deformed background presented here.

\subsection{$\mathfrak r'_{3,0}\oplus \mathbbm{R}$}
In this case we can choose an $R$-matrix which involves generators along spacelike directions ($R_{11}$ in table~\ref{tab:R-matrices})
\begin{equation}
R=J_{12}\wedge p_3 + p_2\wedge p_1\, .
\end{equation}
It preserves 3 bosonic isometries
\begin{equation}
J_{12},\qquad p_0,\qquad p_3,
\end{equation}
and no supercharges.
It is more convenient to use the parameterisation
\begin{equation}
g_a = \exp(\xi J_{12})\cdot \exp(rp_1+x^0p_0+x^3p_3)\cdot \exp(\log z\, D),
\end{equation}
since $\xi$ will be isometric. In the undeformed case
\begin{equation}
ds^2_{\eta=0}=\frac{-(d{x^0})^2+r^2 d\xi^2+dr^2+d{x_3}^2+dz^2}{z^2}+ds_s^2,
\end{equation}
so that $(r,\xi)$ are a radial and an angular coordinate in the $1,2$ plane.
Turning on the deformation parameter we find
\begin{equation}
\begin{aligned}
ds^2&=z^{-6} \left(1+\frac{4 \eta ^2 \left(r^2+1\right)}{z^4}\right)^{-1}\left[dr^2 \left(4 \eta ^2 r^2+z^4\right)+r^2 z^4 d\xi^2-8 \eta ^2 r \, dr d {x_3}+d {x_3}^2 \left(4 \eta
   ^2+z^4\right)\right]\\
&+\frac{dz^2-(d {x^0})^2}{z^2}+ds_s^2
\end{aligned}
\end{equation}
The dilaton and the $B$-field now depend on $r$ and $z$
\begin{equation}
e^\phi = \left( 1+\frac{4 \eta ^2 \left(r^2+1\right)}{z^4}\right)^{-1/2},
\qquad\qquad
B=\frac{2 \eta \,  r\, d\xi \wedge ( dr+r  d {x_3})}{z^4+4 \eta ^2 \left(r^2+1\right)}.
\end{equation}
For the RR-fluxes we find
\begin{equation}
F^{(5)}=(1+*)\frac{4 r \ d{x^0}\wedge d r\wedge d \xi \wedge d  {x_3}\wedge d z}{z \left(z^4+4 \eta ^2 \left(r^2+1\right)\right)},
\qquad
F^{(3)}= \frac{8 \eta  }{z^5} (d {x_3}-r dr)\wedge d{x^0}\wedge dz.
\end{equation}
The sequence of TsT-transformations
\begin{equation}
T(x_3),\ \xi\to \xi+2\eta \tilde x_3,\ T(\tilde x_3),
\qquad\qquad
T(x_1),\ x_2\to x_2-2\eta \tilde x_1,\ T(\tilde x_1),
\end{equation}
(where $r=\sqrt{x_1^2+x_2^2},\ \xi=\arctan(x_1/x_2)$) yields undeformed $AdS_5\times S^5$.

\subsection{$\mathfrak r_{3,-1}\oplus \mathbbm{R}$}
The $R$-matrix ($R_6$ in table~\ref{tab:R-matrices} with $x^1\rightarrow x^2$, $x^3\rightarrow x^1$)
\begin{equation}
R= J_{01}\wedge p_2 + 2 p_0\wedge p_1\,,
\end{equation}
preserves 3 bosonic isometries
\begin{equation}
J_{01},\qquad p_2,\qquad p_3\,,
\end{equation}
and no supercharges.
As before, it is more convenient to parameterise the group element in a different way
\begin{equation}
g_a = \exp(t J_{01})\cdot \exp(\rho p_1+x^2p_2+x^3p_3)\cdot \exp(\log z\, D),
\end{equation}
so that $t$ is an isometry. In the undeformed case we have the spacetime metric
\begin{equation}
ds^2_{\eta=0} =\frac{-\rho ^2 dt^2+d\rho ^2+d {x_2}^2+d {x_3}^2+dz^2}{z^2} + ds_s^2,
\end{equation}
while the defomation gives
\begin{equation}
\begin{aligned}
ds^2 &=z^{-6} \left(1-\frac{4 \eta ^2 \left(\rho
   ^2+4\right)}{z^4}\right)^{-1}\left(-\rho ^2 z^4 dt^2-16 \eta ^2 \rho  d\rho  d {x_2}+d {x_2}^2 \left(z^4-16 \eta ^2\right)+d\rho ^2
   \left(z^4-4 \eta ^2 \rho ^2\right)\right)\\
&+\frac{d {x_3}^2}{z^2}+\frac{dz^2}{z^2}
 + ds_s^2\,.
\end{aligned}
\end{equation}
The dilaton and the $B$-field depend on $\rho$ and $z$
\begin{equation}
e^\phi=\left( 1-\frac{4\eta^2(4+\rho^2)}{z^4} \right)^{-1/2}\,,
\qquad
\qquad
B=\frac{2 \eta\,  \rho\,  dt\wedge (2 d\rho -\rho  d{x_2})}{z^4-4 \eta ^2 \left(4+\rho ^2\right)}\,,
\end{equation}
and the RR-fluxes are
\begin{equation}
F^{(5)}= -(1+*)\frac{4 \rho\,  dt\wedge d\rho \wedge d {x_2}\wedge d {x_3}\wedge dz}{z \left(z^4-4 \eta ^2 \left(4+\rho ^2\right)\right)}\,,
\qquad
F^{(3)}=\frac{8 \eta (2 d {x_2}+\rho  d\rho)\wedge d {x_3}\wedge dz}{z^5}\,.
\end{equation}
We can get back the undeformed $AdS_5\times S^5$ background by applying the sequence of TsT-transformations
\begin{equation}
T(x_2),\ t\to t+2\eta \tilde x_2,\ T(\tilde x_2),
\qquad\qquad
T(x_1),\ x_0\to x_0-4\eta \tilde x_1,\ T(\tilde x_1),
\end{equation}
where $x_1=\rho \cosh t, \ x_0=\rho \sinh t$.

\subsection{$\mathfrak{n}_4$}
Let us consider the $R$-matrix ($R_{15}$ in table~\ref{tab:R-matrices} with $x^1\leftrightarrow x^3$)
\begin{equation}
R=p_1 \wedge p_3 +(p_0+p_1)\wedge (J_{03}+J_{13})
\end{equation}
which preserves the 3 bosonic isometries
\begin{equation}
p_0+p_1\,, \qquad
p_2\,,\qquad
p_3\,,
\end{equation}
and 8 supercharges.
The metric is given by
\begin{equation}
\begin{aligned}
ds^2
&=z^{-6}\left(1-\frac{4 \eta   ^2 \xi_-}{z^4}\right)^{-1}\left[z^4 d {x_3}^2-\eta ^2 (d {x^+})^2-\frac{1}{4} d\xi_-  \left(\eta ^2 \xi_- ^2 d\xi_- +2 d {x^+} \left(z^4-2 \eta ^2 \xi_-
   \right)\right)\right]\\
&+\frac{d {x_2}^2+dz^2}{z^2}+ds_s^2\,,
\end{aligned}
\end{equation}
where we preferred to redefine $\xi_-=2x^--1$.
The dilaton and the $B$-field depend on $\xi_-$ and $z$
\begin{equation}
e^\phi = \left(1-\frac{4 \eta   ^2 \xi_-}{z^4}\right)^{-1/2}\,,
\qquad
\qquad
B=\frac{\eta   (\xi_-  d\xi_- +2 d {x^+})\wedge d {x_3} }{2 \left(z^4-4 \eta ^2 \xi_- \right)}.
\end{equation}
The RR-fluxes are
\begin{equation}
\begin{aligned}
F^{(5)}=(1+*)\frac{d\xi_-\wedge d {x^+}\wedge  d {x_2}\wedge d {x_3}\wedge dz}{z(z^4-4 \eta ^2 \xi_-  )},\qquad
F^{(3)}=\frac{2\eta }{z^5} \left(\xi_-  d\xi_--2 d{x^+}\right)\wedge d {x_2}\wedge d z.
\end{aligned}
\end{equation}
We have checked that this background is self-dual (after field redefinitions) under a TsT-transformation involving $p_0+p_1$ and $p_3$. If we view it as a deformation of $AdS_4$ there are no other bosonic isometries at our disposal, so it appears that this background cannot be generated by (bosonic) TsT-transformations. As remarked earlier, it would be very interesting to understand if it can be generated by applying non-abelian T-duality.

\section{Conclusions}
We have derived the target space geometry of the $\eta$ and $\lambda$-deformed type IIB supercoset string sigma models. With this result we have checked that the $\lambda$-deformation leads to a (type II*) supergravity background, while  in general the  $\eta$-deformation only to a ``generalized'' one in the sense of \cite{Arutyunov:2015mqj,Wulff:2016tju}. When this is the case, the sigma model is expected to be scale invariant but not Weyl invariant, and therefore does not seem to define a consistent string theory. We have identified the (necessary and sufficient) condition for the $\eta$-model to have a \emph{standard} supergravity background as target space. This is translated into an algebraic condition on the $R$-matrix, which we refer to as the \emph{unimodularity condition}.  It imposes strong restrictions on non-abelian $R$-matrices, and in fact all non-abelian $R$-matrices considered in previous works do not lead to supergravity solutions.

We have also analyzed the problem of finding all unimodular $R$-matrices which solve the CYBE for the bosonic subalgebra $\mathfrak{so}(2,4)\oplus\mathfrak{so}(6)\subset\mathfrak{psu}(2,2|4)$. The complete list of rank four non-abelian $R$-matrices for $\mathfrak{so}(2,4)$ has been given and we have showed that the only other non-abelian $R$-matrices in this case have rank six. We have argued that most of these examples should correspond to a sequence of non-commuting TsT-transformations and have verified this explicitly in some cases. It should be possible to understand these deformations in terms of twisted boundary conditions for the string just as in the standard TsT case~\cite{Frolov:2005dj}. There are many similarities between the backgrounds we construct and that of Hashimoto-Itzhaki/Maldacena-Russo \cite{Hashimoto:1999ut,Maldacena:1999mh} and the dual field theories are expected to be certain non-commutative deformations of $\mathcal N=4$ super Yang-Mills, see \cite{Matsumoto:2014gwa} and in particular \cite{vanTongeren:2015uha}. 

Many interesting open questions remain. It would be important to find all possible unimodular $R$-matrices of $\mathfrak{psu}(2,2|4)$ to have a complete list of  Yang-Baxter deformations of $AdS_5\times S^5$ with a string theory interpretation. A question is whether any of them are of the Jordanian type. It is particularly interesting to investigate whether it is possible to have unimodular $R$-matrices that solve the MCYBE rather than the CYBE, to solve one of the puzzles of~\cite{Arutyunov:2015qva}. 
One could also try to give an interpretation to backgrounds generated by non-unimodular $R$-matrices; in many cases one can associate to them a formally T-dual model which does describe a string sigma model, so it is natural to wonder what these backgrounds correspond to. See \cite{Orlando:2016qqu} for some investigations along these lines.
It would be also interesting to clarify if these deformed models have a connection to non-abelian T-duality, in view of the similarities between our unimodularity condition and the tracelessness condition of~\cite{Elitzur:1994ri}.

Our results are also useful to make further progress in the case of the $\lambda$-model. In fact, we have written the NSNS and RR background fields in terms of the Lie algebra operators which are used to define the deformation procedure, and after picking a certain parameterisation for the group element this enables to obtain their explicit form. This method is more efficient, albeit equivalent, to the ones used so far e.g. in~\cite{Arutyunov:2015qva,Borsato:2016zcf,Orlando:2016qqu}. One could then check the proposal of~\cite{Chervonyi:2016ajp} for the background of the $\lambda$-deformed $AdS_3\times S^3\times T^4$ string, and finally derive the one for the $AdS_5\times S^5$ case. 
It would be interesting to understand whether there is room to modify the definition of the $\lambda$-model, hence realising other possible deformations of the string. In fact, in the current status the $\lambda$-model is related through Poisson-Lie T-duality to the $\eta$-model based on the MCYBE, but there is no known counterpart for deformations based on the CYBE.

\section*{Acknowledgements}
We would like to thank Arkady Tseytlin for useful discussions and helpful comments on a first draft of this manuscript.
This work was supported by the ERC
Advanced grant No.~290456.
\appendix

\section{$\mathbbm{Z}_4$-graded superisometry algebras}\label{app:superalgebras}
In this appendix we review some facts about the relevant superalgebras and explain our notation and conventions.
In~\cite{Wulff:2015mwa} it was shown that for all cases of interest here\footnote{We restrict our attention to models with only RR flux since these have certain simplifying features like $\mathbbm{Z}_4$-symmetry.} the superisometry algebra---which admits a $\mathbbm Z_4$-grading that extends the $\mathbbm Z_2$-grading of the bosonic subalgebra---can be written in the same form. The bosonic subalgebra is of the standard symmetric space form
\begin{align}
&[J_{ab},P_c]=2\eta_{c[a}P_{b]}\,,\quad
[P_a,P_b]=\frac12\widehat{\mathcal K}_{ab}{}^{cd}J_{cd}\,,
\nonumber\\
&[J_{ab},J_{cd}]=\eta_{ac}J_{bd}-\eta_{bc}J_{ad}-\eta_{ad}J_{bc}+\eta_{bd}J_{ac}\,.
\label{eq:alg-b}
\end{align}
Here $a,b,c=0,\ldots,9$ and $J_{ab}$ generate Lorentz-transformations and rotations while $P_a$ generate translations. Note that since the space is typically a product of factors $J_{ab}$ is block-diagonal with components mixing different factors absent and this should be taken into account in interpreting the last commutator above. In the case of RR backgrounds, i.e. no NSNS three-form flux, the commutators involving the supercharges take the form (here and in the rest of the paper we specialize to the type IIB case, but the type IIA case works in the same way)
\begin{align}
&[P_a,Q^I_{\hat\alpha}]=-i(Q^J\widehat{\mathcal K}^{JI}\gamma_a)_{\hat\alpha}\,,\qquad
[J_{ab},Q^I_{\hat\alpha}]=-\frac12(Q^I\gamma_{ab})_{\hat\alpha}\,,\qquad (I,J=1,2)
\nonumber\\
&\{Q^1_{\hat\alpha},Q^1_{\hat\beta}\}=\{Q^2_{\hat\alpha},Q^2_{\hat\beta}\}=i\gamma^a_{\hat\alpha\hat\beta}\,P_a\,,\qquad
\{Q^1_{\hat\alpha},Q^2_{\hat\beta}\}=(\gamma^a\widehat{\mathcal K}^{12}\gamma^b)_{\hat\alpha\hat\beta}\,J_{ab}\,.
\label{eq:alg-f}
\end{align}
Here $\hat\alpha=1,\ldots,N$ where $2N$ is the number of supersymmetries preserved by the background. For $AdS_5\times S^5$ ($\mathfrak{psu}(2,2|4)$) $N=16$ and $\gamma^a_{\hat\alpha\hat\beta}$ are the standard $16\times16$ symmetric Weyl blocks or `chiral gamma-matrices' (see for example the appendix of~\cite{Wulff:2013kga}). For $AdS_3\times S^3\times T^4$ ($\mathfrak{psu}(1,1|2)^2$) $N=8$ and for $AdS_2\times S^2\times T^6$ ($\mathfrak{psu}(1,1|2)$) $N=4$ and the gamma-matrices $\gamma^a_{\hat\alpha\hat\beta}$ involve an extra projector to make them $8\times8$ and $4\times4$ respectively. The $\mathbbm Z_4$ automorphism acts as
\begin{equation}
J_{ab}\rightarrow J_{ab}\,,\qquad P_a\rightarrow-P_a\,,\qquad Q^1\rightarrow iQ^1\,,\qquad Q^2\rightarrow-iQ^2\,.
\end{equation}
We introduce projectors that split the generators $T_A=\{P_a,\,J_{ab},\,Q^I_{\hat\alpha}\}$ according to their $\mathbbm Z_4$-grading as follows 
\begin{equation}
P^{(0)}(T_A)=J_{ab}\,,\qquad P^{(1)}(T_A)=Q^1_{\hat\alpha}\,,\qquad P^{(2)}(T_A)=P_a\,,\qquad P^{(3)}(T_A)=Q^2_{\hat\alpha}\,.
\end{equation}
Finally $\widehat{\mathcal K}^{AB}$ appearing on the right-hand-side in (\ref{eq:alg-b}) and (\ref{eq:alg-f}) is the inverse of the Lie algebra metric defined by the supertrace\footnote{Note that our definition of $\mathcal K$ differs by a factor of $i$ compared to the definition used in~\cite{Wulff:2015mwa}.}
\begin{equation}
\mathrm{Str}(T_AT_B)=\mathcal K_{AB}\,,\qquad T_A=\{P_a,\,J_{ab},\,Q^I_{\hat\alpha}\}\,,
\end{equation}
e.g. 
\begin{equation}
\frac12\widehat{\mathcal K}_{ab}{}^{ef}\mathcal K_{ef,cd}=2\eta_{a[c}\eta_{d]b}\,.
\end{equation}
It can be expressed in terms of the geometry and fluxes of the corresponding symmetric space supergravity background as
\begin{equation}
\widehat{\mathcal K}^{ab}=\eta^{ab}\,,\qquad\widehat{\mathcal K}_{ab}{}^{cd}=-\underline R_{ab}{}^{cd}\,,\qquad
\widehat{\mathcal K}^{\hat\alpha I\hat\beta J}=\frac{i}{8}\underline{\mathcal S}^{\hat\alpha I\hat\beta J}\,,
\end{equation}
where $\underline R_{ab}{}^{cd}$ and $\underline{\mathcal S}^{IJ}$ are the Riemann curvature and RR field strength bispinor respectively.\footnote{The curvature of $AdS$ is $R_{ab}{}^{cd}=2\delta_{[a}^c\delta_{b]}^d$ while that of the sphere is $R_{ab}{}^{cd}=-2\delta_{[a}^c\delta_{b]}^d$ in our conventions. The RR flux takes the form
$$
AdS_n\times S^n\times T^{10-2n}:\qquad\underline{\mathcal S}^{\hat\alpha I\hat\beta J}=-4i(\sigma^2)^{IJ}(\mathcal P\gamma^{01234})^{\hat\alpha\hat\beta}\,,
$$
where the projector $\mathcal P$, with $Q^I=Q^I\mathcal P$, is given by $1$ for $n=5$, $\frac12(1+\gamma^{6789})$ for $n=3$ and $\frac12(1+\gamma^{6789})\frac12(1+\gamma^{4568})$ for $n=2$.
%
} Let us also note the relation
\begin{equation}
\widehat{\mathcal K}_{ab}{}^{cd}(\mathcal K^{12}\gamma_{cd})_{\hat\alpha\hat\beta}=8(\gamma_{[a}\widehat{\mathcal K}^{12}\gamma_{b]})_{\hat\alpha\hat\beta}\,.
\end{equation}

Finally for operators acting on the Lie algebra (i.e. endomorphisms) $\mathcal M:\,\mathfrak g\rightarrow\mathfrak g$ we define its components in the following way
\begin{equation}
\mathcal M(T_C)=T_D\mathcal M^D{}_C\,.
\end{equation}
The transpose operator is defined with respect to the supertrace by
\begin{equation}
\mathrm{Str}(T_A\mathcal M(T_B))=\mathrm{Str}(\mathcal M^T(T_A)T_B)\,,
\end{equation}
or
\begin{equation}
\mathcal M_{AB}=(-1)^{AB}(\mathcal M^T)_{BA}
\qquad\mathcal M_{AB}=\mathcal K_{AC}\mathcal M^C{}_B
\label{eq:MT-comp}
\end{equation}
e.g.
\begin{equation}
(\mathcal{M}^T)_{a\hat\beta1}
=
\mathcal K_{\hat\beta1\hat\gamma2}M^{\hat\gamma2}{}_a\,,
\qquad
%
(\mathcal M^T)_{a,bc}=\frac12\mathcal K_{bc,de}\mathcal M^{de}{}_a\,.
\end{equation}
The supertrace of the Lie algebra operator $\mathcal M$ is given by
\begin{equation}\label{eq:Str}
\mathrm{Str}(\mathcal M)=(-1)^A\mathcal M^A{}_A=\widehat{\mathcal  K}^{AB}\mathrm{Str}(T_A\mathcal MT_B)\,.
\end{equation}
When we need to raise indices with $\widehat{\mathcal  K}^{AB}$ we use the convention 
\begin{equation}
\mathcal M^A=\mathcal M_B\widehat{\mathcal  K}^{BA}\,.
\end{equation}
To conclude, when writing generic commutation relations we write
\begin{equation}
[T_A,T_B]=f^C{}_{AB}T_C\,.
\end{equation}

\section{Useful results for the deformed models}\label{app:def}
In this appendix we collect some useful identities and expressions to obtain the results presented in the main text.
In the two deformed models, we can relate $\mathcal O^T_\pm$ and $\mathcal O_\pm$ by
\begin{equation}
\mathbf\lambda-\mbox{\bf model}:\qquad \mathcal O_-^T=\Ad_g^{-1}\mathcal O_+,\qquad
\mathbf\eta-\mbox{\bf model}:\qquad 
\mathcal O_-^T\hat d^T=\hat d^T\mathcal O_+\,.
\end{equation}
Using the definitions of $\mathcal O_\pm$, we can express $M$ defined in~\eqref{eq:def-M} in terms of $\mathcal O_\pm$ and projectors only 
\begin{equation}
\begin{split}
\mathbf\lambda-\mbox{\bf model}:&\qquad M
=-\Omega^T+(\mathcal O_+^T)^{-1}(1-\Omega\Omega^T)
=-\Omega^T+(1-\lambda^{-4})(\mathcal O_+^T)^{-1}P^{(2)}\,,
\\
\mathbf\eta-\mbox{\bf model}:&\qquad 
M=
\mathcal O_-^{-1}(\mathcal O_-+2\eta R_g\hat dP^{(2)})
=
1-2P^{(2)}+2\mathcal O_-^{-1}P^{(2)}\,,
\end{split}
\label{eq:M}
\end{equation}
which is useful to  prove 
\begin{equation}
\begin{split}
\mathbf\lambda-\mbox{\bf model}:&\qquad 
\Ad_h^{-1}P^{(2)}
=\mathcal O_+(1+\Omega(\mathcal O_+^T)^{-1})P^{(2)}
=P^{(2)}(1+(\mathcal O_+^T)^{-1}\Omega)\mathcal O_+\,,
\\
\mathbf\eta-\mbox{\bf model}:&\qquad 
\Ad_h^{-1}P^{(2)}
=\mathcal O_+(2P^{(2)}-1)\mathcal O_-^{-1}P^{(2)}\,.
\end{split}
\label{eq:identity2}
\end{equation}
Note that using the expression for $M$ we can express $A_-$ in terms of $A_+$ as
\begin{equation}
A_-=MA_+
=
\left\{\begin{array}{c}
A_++(M-1)A_+^{(2)}\\	
-\Omega^TA_++(M+\lambda^{-2})A_+^{(2)}
\end{array}
\right.\,.
\label{eq:Apm-rel}
\end{equation}
The rest of this appendix is devoted to the two deformed models separately.

\subsection{$\lambda$-model}\label{app:lambda}
The expressions for $dA_\pm$ in~\eqref{eq:dAp-lambda},\eqref{eq:dAm-lambda} can be rewritten as
\begin{align}
dA_+
=&
-\frac12\{A_+,A_+\}
-\frac12(1-\lambda^{-4})
\mathcal O_+^{-1}
(
\{A_+^{(2)},A_+^{(2)}\}
-\lambda^2\{A_+^{(1)},A_+^{(1)}\}
+2\lambda\{A_+^{(2)},A_+^{(3)}\}
)\,,
\label{eq:dA-plus-lambda}
\\
dA_-=&
\frac12\{A_-,A_-\}
+\frac12(1-\lambda^{-4})
(\mathcal O_+^T)^{-1}
(
\{A_-^{(2)},A_-^{(2)}\}
-\lambda^2\{A_-^{(3)},A_-^{(3)}\}
+2\lambda\{A_-^{(2)},A_-^{(1)}\}
)\,,
\label{eq:dA-minus-lambda}
\end{align}
if we use
\begin{equation}
\Omega^T\{X,X\}-\{\Omega^TX,\Omega^TX\}
=
(1-\lambda^{-4})
(
\{X^{(2)},X^{(2)}\}
-\lambda^2\{X^{(1)},X^{(1)}\}
+2\lambda\{X^{(2)},X^{(3)}\}
),
\end{equation}
for $X\in\mathfrak g$, and the same for $\Omega$ but with $X^{(1)}$ and $X^{(3)}$ interchanged.

To calculate the component $T^{\hat\alpha1}$ of the torsion, we first need to compute the Lorentz-transformed spin-connection $\Ad_h A^{(0)}_++dhh^{-1}$.
We do this by taking the exterior derivative of both sides of the relation $E^{(2)}=\Ad_h A_-^{(2)}$, from which we find the equation
\begin{align}
0=&
\{\Ad_hA_+^{(0)}+dhh^{-1}-A_+^{(0)},E^{(2)}\}
+\lambda(1-\lambda^{-4})P^{(2)}\Ad_h(\mathcal O_+^T)^{-1}\Ad_h^{-1}\{E^{(2)},E^{(1)}\}%
\nonumber\\
{}&
+\{E^{(1)},\Ad_hP^{(1)}ME^{(2)}\}%
-i\lambda\{E^{(3)},P^{(3)}ME^{(2)}\}
-i\lambda P^{(2)}M^T\{E^{(2)},E^{(3)}\}
\nonumber\\
{}&
-\{\Ad_hP^{(0)}ME^{(2)},E^{(2)}\}
-\frac12\Ad_h\{P^{(1)}ME^{(2)},P^{(1)}ME^{(2)}\}%
-\frac12\lambda^2\{P^{(3)}ME^{(2)},P^{(3)}ME^{(2)}\}
\nonumber\\
{}&
-\frac12(1-\lambda^{-4})P^{(2)}\Ad_h(\mathcal O_+^T)^{-1}\Ad_h^{-1}
(
\{E^{(2)},E^{(2)}\}
+2\lambda\{E^{(2)},\Ad_hP^{(1)}ME^{(2)}\}
)
\nonumber\\
{}&
-\frac12P^{(2)}M^T\{E^{(2)},E^{(2)}\}
-\lambda^2 P^{(2)}M^T\{E^{(2)},P^{(3)}ME^{(2)}\}
\,,
\end{align}
where we used (\ref{eq:dE2-lambda}) and (\ref{eq:dA-minus-lambda}). This equation determines $\Ad_hA_+^{(0)}+dhh^{-1}$ completely: this is obvious for the terms involving fermionic vielbeins, while for the terms involving $E^a$ it follows from symmetry in the same way that the condition $T_{ab}{}^c=0$ determines the spin connection $\Omega_{ab}{}^c$. Using the algebra (\ref{eq:alg-b}), (\ref{eq:alg-f}) as well as (\ref{eq:identity2}) the result is
\begin{align}
[dhh^{-1}+\Ad_hA_+]_{ab}
=
-\Omega_{ab}
+\frac12E^cH_{abc}
+2i(E^1\gamma_{[a})_{\hat\alpha}(\Ad_hM)^{\hat\alpha1}{}_{b]}
\,.
\label{eq:dh-lambda}
\end{align}
Here we have used the fact, which will be proven below, that the expression that we find
\begin{equation}
H_{abc}=3[\Ad_hM]_{[ab,c]}+3iM^{\hat\alpha1}{}_{[a}[\Ad_h]_{b|d|}\gamma^d_{\hat\alpha\hat\beta}M^{\hat\beta1}{}_{c]}\,,
\label{eq:Habc-lambda2}
\end{equation}
is equivalent to the one in (\ref{eq:Habc-lambda}).
In fact, if we calculate $H=dB$ using the first definition for $B$ in~\eqref{eq:B-lambda}
we find
\begin{align}
H=&
dB
=
\frac13(1-\lambda^{-4})^{-1}\big(\mathrm{Str}(\Omega A_-\wedge\Omega A_-\wedge\Omega A_-)-\mathrm{Str}(A_-\wedge A_-\wedge A_-)\big)
\nonumber\\
&{}
-\frac12(1-\lambda^{-4})^{-1}\mathrm{Str}(A_+\wedge(\Omega\{A_-,A_-\}-\{\Omega A_-,\Omega A_-\}))
\nonumber\\
=&
-\mathrm{Str}((A_+^{(0)}+A_-^{(0)})\wedge A_-^{(2)}\wedge A_-^{(2)})
+\lambda^2\mathrm{Str}(A_+^{(2)}\wedge A_-^{(3)}\wedge A_-^{(3)})
\nonumber\\
&{}
-\frac12\mathrm{Str}(A_-^{(2)}\wedge\{A_-^{(1)},A_-^{(1)}+2\lambda A_+^{(1)}\})
\nonumber\\
=&
\mathrm{Str}(E^{(2)}\wedge E^{(1)}\wedge E^{(1)})
-\mathrm{Str}(E^{(2)}\wedge E^{(3)}\wedge E^{(3)})
-\mathrm{Str}(P^{(0)}\Ad_hME^{(2)}\wedge E^{(2)}\wedge E^{(2)})
\nonumber\\
&{}
-\mathrm{Str}(E^{(2)}\wedge P^{(1)}\Ad_hME^{(2)}\wedge P^{(1)}\Ad_hME^{(2)})
\nonumber\\
=&
-\frac{i}{2}E^a\,E^1\gamma_aE^1
+\frac{i}{2}E^a\,E^2\gamma_aE^2
+\frac{1}{3!}E^cE^bE^aH_{abc}\,,
\end{align}
with $H_{abc}$ given by (\ref{eq:Habc-lambda2}). 
On the other hand, if we start from $B$ given in the second line of~\eqref{eq:B-lambda}, we find a result which is mapped to the previous one by the replacements $A_-\leftrightarrow A_+$, $\Omega\leftrightarrow\Omega^T$ and $A^{(3)}\leftrightarrow A^{(1)}$. This leads to the same form of $H$ except now with $H_{abc}$ given by (\ref{eq:Habc-lambda}), which proves the equivalence of the two expressions.
Let us also remark that this computation shows that the NSNS three-form superfield $H=dB$ satisfies the correct superspace constraints.

In order to check that the dilatinos in~\eqref{eq:chi2-lambda},\eqref{eq:chi1-lambda} are in fact the spinor derivatives of the dilaton $\phi$, we start from~\eqref{eq:e-2phi-lambda} and compute
\begin{align}
d\phi=&
-\frac12\mathrm{STr}(\mathcal O_-^{-1}\Ad_g^{-1}d\Ad_g)
=
-\frac12\widehat{\mathcal K}^{AB}\mathrm{STr}(T_A\mathcal O_-^{-1}[g^{-1}dg,T_B])
\nonumber\\
=&
-\frac12\widehat{\mathcal K}^{AB}\mathrm{STr}(T_A\mathcal O_-^{-1}[\mathcal O_-A_-,T_B])
\nonumber\\
=&
-\frac12\widehat{\mathcal K}^{AB}\mathrm{STr}(T_A\mathcal O_-^{-1}[A_-,T_B])
+\frac12\widehat{\mathcal K}^{AB}\mathrm{STr}(T_A\mathcal O_-^{-1}\Ad_g^{-1}[\Omega A_-,\Ad_gT_B])
\nonumber\\
=&
-\frac12\widehat{\mathcal K}^{AB}\mathrm{STr}(T_A\mathcal O_-^{-1}[A_-,T_B])
+\frac12\widehat{\mathcal K}^{AB}\mathrm{STr}(T_A\Ad_g\mathcal O_-^{-1}\Ad_g^{-1}[\Omega A_-,T_B])
\nonumber\\
=&
-\frac12\widehat{\mathcal K}^{AB}\mathrm{STr}(T_A\mathcal O_-^{-1}[A_-,T_B])
+\frac12\widehat{\mathcal K}^{AB}\mathrm{STr}(T_A\Ad_g(\mathcal O_+^T)^{-1}[\Omega A_-,T_B])
\nonumber\\
=&
-\frac12\widehat{\mathcal K}^{AB}\mathrm{STr}(T_A\mathcal O_-^{-1}[A_-,T_B])
+\frac12\widehat{\mathcal K}^{AB}\mathrm{STr}(T_A\Omega\mathcal O_-^{-1}Ad_g^{-1}[\Omega A_-,T_B])
\nonumber\\
=&
-\frac12\widehat{\mathcal K}^{AB}\mathrm{STr}(T_A\mathcal O_-^{-1}[A_-,T_B])
+\frac12\widehat{\mathcal K}^{AB}\mathrm{STr}(T_A\Omega\mathcal O_-^{-1}\Omega^T[\Omega A_-,T_B])
\nonumber\\
&{}
+\frac12(1-\lambda^{-4})\lambda^2\widehat{\mathcal K}^{AB}\mathrm{STr}(T_A\Omega\mathcal O_-^{-1}P^{(2)}[\Omega A_-,T_B])\,,
\end{align}
where we used~\eqref{eq:Str} and in the last step we inserted $1=1-\Omega\Omega^T+\Omega\Omega^T=(1-\lambda^{-4})P^{(2)}+\Omega\Omega^T$. It is easy to see that the $A_-^{(0)}$-terms cancel, as they must since they transform as a connection. 

\subsection{$\eta$-model}\label{app:eta}
The expressions for $dA_\pm$ in~\eqref{eq:dAp-eta}, \eqref{eq:dAm-eta} can be rewritten as
\begin{align}
dA_+=\,&
\frac12\{A_+,A_+\}
-\frac12c\eta^2\{\hat d^TA_+,\hat d^TA_+\}
+(\mathcal O_+^{-1}-1)
\Big(
4\{A_+^{(2)},A_+^{(3)}\}
+\hat\eta^2\{A_+^{(1)},A_+^{(1)}\}
\Big)
\nonumber\\
&{}
+\eta\mathcal O_+^{-1}R_g\{A_+^{(2)},\hat d^TA_+^{(2)}\}\,,
\label{eq:dA-plus-eta}
\\
dA_-=\,&
\frac12\{A_-,A_-\}
-\frac12c\eta^2\{\hat dA_-,\hat dA_-\}
+(\mathcal O_-^{-1}-1)
\Big(
4\{A_-^{(2)},A_-^{(1)}\}
+\hat\eta^2\{A_-^{(3)},A_-^{(3)}\}
\Big)
\nonumber\\
&{}
-\eta\mathcal O_-^{-1}R_g\{A_-^{(2)},\hat dA_-^{(2)}\}\,,
\label{eq:dA-minus-eta}
\end{align}
where we have rewritten e.g. the last term in the expression for $dA_+$ as
\begin{equation}
\begin{aligned}
&\eta\mathcal O_+^{-1}R_g\{A_+^{(2)},\hat d^TA_+^{(2)}\}\\
&+(1-\mathcal O_+^{-1})
\Big(
\frac12\{A_+,A_+\}
-\frac12c\eta^2\{\hat d^TA_+,\hat d^TA_+\}
-4\{A_+^{(2)},A_+^{(3)}\}
-\hat\eta^2\{A_+^{(1)},A_+^{(1)}\}
\Big)\,.
\end{aligned}
\end{equation}

As in the case of the $\lambda$-model, to calculate the component $T^{\hat\alpha 1}$ of the torsion we must first find the Lorentz-transformed spin connection $\Ad_hA_+^{(0)}-dhh^{-1}$ (note the difference in sign between the two models). We use the same method explained in the previous subsection and we find
\begin{equation}
[\Ad_hA_+^{(0)}-dhh^{-1}]_{ab}
=
\Omega_{ab}
-\frac12E^cH_{abc}
+2i\hat\eta(\gamma_{[a}E^1)_{\hat\alpha}[\Ad_hM]^{\hat\alpha1}{}_{b]}\,,
%
\label{eq:dh-eta}
\end{equation}
where we write the components of $H_{abc}$ as
\begin{equation}
H_{abc}=3[\Ad_hM]_{[ab,c]}-3i\hat\eta^2[\Ad_h]_{[a|d|}M^{\hat\alpha1}{}_b\gamma^d_{\hat\alpha\hat\beta}M^{\hat\beta1}{}_{c]}\,.
\label{eq:Habc-eta2}
\end{equation}
This expression is equivalent to the one in (\ref{eq:Habc-eta}), which is easy to verify by a calculation similar to the one performed for the $\lambda$-model: the $B$-field written as in the first way of~\eqref{eq:B-eta} leads to $H_{abc}$ of the form (\ref{eq:Habc-eta}), while the second way leads to the form in (\ref{eq:Habc-eta2}). The same calculation also shows that the remaining components of the superform $H$ satisfy the standard supergravity constraints.

If we take~\eqref{eq:e-2phi-eta} as the definition of the dilaton in the case of the $\eta$-model we find
\begin{align}
d\phi
=&\,
-\frac12\eta\widehat{\mathcal K}^{AB}\mathrm{STr}(T_A\hat d^T\mathcal O_+^{-1}R_g[g^{-1}dg,T_B])
+\frac12\eta\widehat{\mathcal K}^{AB}\mathrm{STr}(T_AR_g\hat d^T\mathcal O_+^{-1}[g^{-1}dg,T_B])
\nonumber\\
=&\,
-\frac12\eta\widehat{\mathcal K}^{AB}\mathrm{STr}(T_A\hat d^T\mathcal O_+^{-1}R_g[A_+,T_B])
-\frac12\widehat{\mathcal K}^{AB}\mathrm{STr}(T_A\mathcal O_+^{-1}[A_+,T_B])
\nonumber\\
&{}
-\frac12\eta\widehat{\mathcal K}^{AB}\mathrm{STr}(T_A\mathcal O_+^{-1}[R_g\hat d^TA_+,T_B])
-\frac12\eta^2\widehat{\mathcal K}^{AB}\mathrm{STr}(T_A\hat d^T\mathcal O_+^{-1}R_g[R_g\hat d^TA_+,T_B])
\nonumber\\
=&\,
-\frac12\widehat{\mathcal K}^{AB}\mathrm{STr}(T_A\mathcal O_+^{-1}[A_+,T_B])
+\frac12c\eta^2\widehat{\mathcal K}^{AB}\mathrm{STr}(T_A\hat d^T\mathcal O_+^{-1}[\hat d^TA_+,T_B])
\nonumber\\
&{}
-\frac12\eta\widehat{\mathcal K}^{AB}\mathrm{STr}(T_A\hat d^T\mathcal O_+^{-1}R_g[A_+,T_B])
-\frac12\eta\widehat{\mathcal K}^{AB}\mathrm{STr}(T_A\mathcal O_+^{-1}R_g[\hat d^TA_+,T_B])
\nonumber\\
&{}
+\frac12\eta\widehat{\mathcal K}^{AB}\mathrm{STr}(T_AR_g[\hat d^TA_+,T_B])\,,
\label{eq:dphi-eta}
\end{align}
where we used the (M)CYBE (\ref{eq:YBE}) in the last step. It is again easy to verify that the $A^{(0)}$-terms cancel, as they must. 

Using (\ref{eq:dalpha1phi}) and (\ref{eq:dalpha2phi}) and (\ref{eq:dh-eta}), the explicit result for the vector $K^a$ in~\eqref{eq:Ka-dchi} is
\begin{align}
K^a=&
\frac{i}{32}\eta(\gamma^a)^{\hat\alpha\hat\beta}
\widehat{\mathcal K}^{AB}\mathrm{STr}\Big([T_A,RT_B]\Ad_g\big([(1-\eta R_g)\Ad_h^{-1}Q^1_{\hat\alpha},\Ad_h^{-1}Q^1_{\hat\beta}]+\hat\eta^{-2}[(1+\eta R_g)Q^2_{\hat\alpha},Q^2_{\hat\beta}]\big)\Big)
\nonumber\\
&{}
+\mbox{fermions}
\nonumber\\
=&
-\frac{\eta}{2}[\hat\eta^{-2}+\Ad_h]^a{}_b\widehat{\mathcal K}^{AB}\mathrm{STr}([T_A,RT_B]gP_bg^{-1})
\nonumber\\
&{}
-\frac{\eta^2}{32}
\big(
\hat\eta^{-2}(\gamma^a\gamma^c)^{\hat\alpha}{}_{\hat\beta}[R_g]^{\hat\beta2}{}_{\hat\alpha2}
-[\Ad_h]^a{}_b(\gamma^b\gamma^c)^{\hat\alpha}{}_{\hat\beta}[R_g]^{\hat\beta1}{}_{\hat\alpha1}
\big)
\widehat{\mathcal K}^{AB}\mathrm{STr}([T_A,RT_B]gP_cg^{-1})
\nonumber\\
&{}
-\frac{i\eta^2}{32}
\big(
[\Ad_h]^a{}_b(\gamma^b\gamma^cK^{12}\gamma^d)^{\hat\alpha}{}_{\hat\beta}[R_g]^{\hat\beta2}{}_{\hat\alpha1}
\nonumber \\
&\qquad\qquad\qquad
-\hat\eta^{-2}(\gamma^a\gamma^cK^{21}\gamma^d)^{\hat\alpha}{}_{\hat\beta}[R_g]^{\hat\beta1}{}_{\hat\alpha2}
\big)
\widehat{\mathcal K}^{AB}\mathrm{STr}([T_A,RT_B]gJ_{cd}g^{-1})
\nonumber\\
&{}
+\mbox{fermions}\,.
\label{eq:Ka}
\end{align}

\bibliographystyle{nb}
\bibliography{biblio}{}

\begin{thebibliography}{10}
\ifx\href\asklfhas\newcommand{\href}[2]{#2}\fi
\ifx\arxivref\asklfhas\newcommand{\arxivref}[2]{\href{http://arxiv.org/abs/#1}{#2}}\fi
\ifx\doiref\asklfhas\newcommand{\doiref}[2]{\href{http://dx.doi.org/#1}{#2}}\fi
\raggedright
\small
\parskip 0pt

\bibitem{Bena:2003wd}
I.~Bena, J.~Polchinski and R.~Roiban,
\textit{``{Hidden symmetries of the AdS$_5 \times$S$^5$ superstring}''},
\textsf{\doiref{10.1103/PhysRevD.69.046002}{Phys.Rev.~D69,~046002~(2004)}},
\texttt{\arxivref{hep-th/0305116}{hep-th/0305116}}.

\bibitem{Arutyunov:2009ga}
G.~Arutyunov and S.~Frolov,
\textit{``{Foundations of the AdS$_5 \times$S$^5$ Superstring. Part I}''},
\textsf{\doiref{10.1088/1751-8113/42/25/254003}{J.Phys.~A42,~254003~(2009)}},
\texttt{\arxivref{0901.4937}{arxiv:0901.4937}}.

\bibitem{Wulff:2014kja}
L.~Wulff,
\textit{``{Superisometries and integrability of superstrings}''},
\textsf{\doiref{10.1007/JHEP05(2014)115}{JHEP~1405,~115~(2014)}},
\texttt{\arxivref{1402.3122}{arxiv:1402.3122}}.

\bibitem{Wulff:2015mwa}
L.~Wulff,
\textit{``{On integrability of strings on symmetric spaces}''},
\textsf{\doiref{10.1007/JHEP09(2015)115}{JHEP~1509,~115~(2015)}},
\texttt{\arxivref{1505.03525}{arxiv:1505.03525}}.

\bibitem{Delduc:2013qra}
F.~Delduc, M.~Magro and B.~Vicedo,
\textit{``{An integrable deformation of the AdS$_5 \times$S$^5$ superstring
  action}''},
\textsf{\doiref{10.1103/PhysRevLett.112.051601}{Phys.Rev.Lett.~112,~051601~(2014)}},
\texttt{\arxivref{1309.5850}{arxiv:1309.5850}}.

\bibitem{Hollowood:2014qma}
T.~J.~Hollowood, J.~L.~Miramontes and D.~M.~Schmidtt,
\textit{``{An Integrable Deformation of the AdS$_5 \times$S$^5$
  Superstring}''},
\textsf{\doiref{10.1088/1751-8113/47/49/495402}{J.Phys.~A47,~495402~(2014)}},
\texttt{\arxivref{1409.1538}{arxiv:1409.1538}}.

\bibitem{Cherednik:1981df}
I.~Cherednik,
\textit{``{Relativistically Invariant Quasiclassical Limits of Integrable
  Two-dimensional Quantum Models}''},
\textsf{\doiref{10.1007/BF01086395}{Theor.Math.Phys.~47,~422~(1981)}}.

\bibitem{Klimcik:2002zj}
C.~Klimcik,
\textit{``{Yang-Baxter sigma models and dS/AdS T duality}''},
\textsf{\doiref{10.1088/1126-6708/2002/12/051}{JHEP~0212,~051~(2002)}},
\texttt{\arxivref{hep-th/0210095}{hep-th/0210095}}.

\bibitem{Klimcik:2008eq}
C.~Klimcik,
\textit{``{On integrability of the Yang-Baxter sigma-model}''},
\textsf{\doiref{10.1063/1.3116242}{J.Math.Phys.~50,~043508~(2009)}},
\texttt{\arxivref{0802.3518}{arxiv:0802.3518}}.

\bibitem{Delduc:2013fga}
F.~Delduc, M.~Magro and B.~Vicedo,
\textit{``{On classical $q$-deformations of integrable sigma-models}''},
\textsf{\doiref{10.1007/JHEP11(2013)192}{JHEP~1311,~192~(2013)}},
\texttt{\arxivref{1308.3581}{arxiv:1308.3581}}.

\bibitem{Sfetsos:2013wia}
K.~Sfetsos,
\textit{``{Integrable interpolations: From exact CFTs to non-Abelian
  T-duals}''},
\textsf{\doiref{10.1016/j.nuclphysb.2014.01.004}{Nucl.Phys.~B880,~225~(2014)}},
\texttt{\arxivref{1312.4560}{arxiv:1312.4560}}.

\bibitem{Hollowood:2014rla}
T.~J.~Hollowood, J.~L.~Miramontes and D.~M.~Schmidtt,
\textit{``{Integrable Deformations of Strings on Symmetric Spaces}''},
\textsf{\doiref{10.1007/JHEP11(2014)009}{JHEP~1411,~009~(2014)}},
\texttt{\arxivref{1407.2840}{arxiv:1407.2840}}.

\bibitem{Tseytlin:1993hm}
A.~A.~Tseytlin,
\textit{``{On A 'Universal' class of WZW type conformal models}''},
\textsf{\doiref{10.1016/0550-3213(94)90243-7}{Nucl.~Phys.~B418,~173~(1994)}},
\texttt{\arxivref{hep-th/9311062}{hep-th/9311062}}.

\bibitem{Delduc:2014kha}
F.~Delduc, M.~Magro and B.~Vicedo,
\textit{``{Derivation of the action and symmetries of the $q$-deformed AdS$_5
  \times$S$^5$ superstring}''},
\textsf{\doiref{10.1007/JHEP10(2014)132}{JHEP~1410,~132~(2014)}},
\texttt{\arxivref{1406.6286}{arxiv:1406.6286}}.

\bibitem{Hollowood:2015dpa}
T.~J.~Hollowood, J.~L.~Miramontes and D.~M.~Schmidtt,
\textit{``{S-Matrices and Quantum Group Symmetry of k-Deformed Sigma
  Models}''},
\texttt{\arxivref{1506.06601}{arxiv:1506.06601}}.

\bibitem{Klimcik:1995ux}
C.~Klimcik and P.~Severa,
\textit{``{Dual nonAbelian duality and the Drinfeld double}''},
\textsf{\doiref{10.1016/0370-2693(95)00451-P}{Phys.Lett.~B351,~455~(1995)}},
\texttt{\arxivref{hep-th/9502122}{hep-th/9502122}}.

\bibitem{Klimcik:1995dy}
C.~Klimcik and P.~Severa,
\textit{``{Poisson-Lie T duality and loop groups of Drinfeld doubles}''},
\textsf{\doiref{10.1016/0370-2693(96)00025-1}{Phys.Lett.~B372,~65~(1996)}},
\texttt{\arxivref{hep-th/9512040}{hep-th/9512040}}.

\bibitem{Vicedo:2015pna}
B.~Vicedo,
\textit{``{Deformed integrable $\sigma$-models, classical $R$-matrices and
  classical exchange algebra on Drinfel'd doubles}''},
\texttt{\arxivref{1504.06303}{arxiv:1504.06303}}.

\bibitem{Hoare:2015gda}
B.~Hoare and A.~Tseytlin,
\textit{``{On integrable deformations of superstring sigma models related to
  AdS$_n \times$S$^n$ supercosets}''},
\texttt{\arxivref{1504.07213}{arxiv:1504.07213}}.

\bibitem{Klimcik:2015gba}
C.~Klimcik,
\textit{``{$\eta$ and $\lambda$ deformations as ${\cal E}$-models}''},
\textsf{\doiref{10.1016/j.nuclphysb.2015.09.011}{Nucl.~Phys.~B900,~259~(2015)}},
\texttt{\arxivref{1508.05832}{arxiv:1508.05832}}.

\bibitem{Arutyunov:2013ega}
G.~Arutyunov, R.~Borsato and S.~Frolov,
\textit{``{S-matrix for strings on $\eta$-deformed AdS$_{5} \times$S$^5$}''},
\textsf{\doiref{10.1007/JHEP04(2014)002}{JHEP~1404,~002~(2014)}},
\texttt{\arxivref{1312.3542}{arxiv:1312.3542}}.

\bibitem{Arutyunov:2015qva}
G.~Arutyunov, R.~Borsato and S.~Frolov,
\textit{``{Puzzles of $\eta$-deformed AdS$_5 \times$S$^5$}''},
\texttt{\arxivref{1507.04239}{arxiv:1507.04239}}.

\bibitem{Hoare:2014pna}
B.~Hoare, R.~Roiban and A.~Tseytlin,
\textit{``{On deformations of AdS$_n \times$S$^n$ supercosets}''},
\textsf{\doiref{10.1007/JHEP06(2014)002}{JHEP~1406,~002~(2014)}},
\texttt{\arxivref{1403.5517}{arxiv:1403.5517}}.

\bibitem{Borsato:2016hud}
R.~Borsato,
\textit{``{Integrable strings for AdS/CFT}''},
\texttt{\arxivref{1605.03173}{arxiv:1605.03173}},
\href{http://inspirehep.net/record/1456973/files/arXiv:1605.03173.pdf}{\texttt{http://inspirehep.net/record/1456973/files/arXiv:1605.03173.pdf}}.

\bibitem{Arutyunov:2015mqj}
G.~Arutyunov, S.~Frolov, B.~Hoare, R.~Roiban and A.~A.~Tseytlin,
\textit{``{Scale invariance of the $\eta$-deformed $AdS_5\times S^5$
  superstring, T-duality and modified type II equations}''},
\textsf{\doiref{10.1016/j.nuclphysb.2015.12.012}{Nucl.~Phys.~B903,~262~(2016)}},
\texttt{\arxivref{1511.05795}{arxiv:1511.05795}}.

\bibitem{Borsato:2016zcf}
R.~Borsato, A.~A.~Tseytlin and L.~Wulff,
\textit{``{Supergravity background of $\lambda$-deformed model for AdS$_2
  \times$ S$^2$ supercoset}''},
\textsf{\doiref{10.1016/j.nuclphysb.2016.02.018}{Nucl.~Phys.~B905,~264~(2016)}},
\texttt{\arxivref{1601.08192}{arxiv:1601.08192}}.

\bibitem{Chervonyi:2016ajp}
Y.~Chervonyi and O.~Lunin,
\textit{``{Supergravity background of the $\lambda$-deformed AdS$_3$ x S$^3$
  supercoset}''},
\textsf{\doiref{10.1016/j.nuclphysb.2016.07.023}{Nucl.~Phys.~B910,~685~(2016)}},
\texttt{\arxivref{1606.00394}{arxiv:1606.00394}}.

\bibitem{Sfetsos:2014cea}
K.~Sfetsos and D.~C.~Thompson,
\textit{``{Spacetimes for $\lambda$-deformations}''},
\textsf{\doiref{10.1007/JHEP12(2014)164}{JHEP~1412,~164~(2014)}},
\texttt{\arxivref{1410.1886}{arxiv:1410.1886}}.

\bibitem{Demulder:2015lva}
S.~Demulder, K.~Sfetsos and D.~C.~Thompson,
\textit{``{Integrable $\lambda$-deformations: Squashing Coset CFTs and
  $AdS_5\times S^5$}''},
\textsf{\doiref{10.1007/JHEP07(2015)019}{JHEP~1507,~019~(2015)}},
\texttt{\arxivref{1504.02781}{arxiv:1504.02781}}.

\bibitem{Wulff:2016tju}
L.~Wulff and A.~A.~Tseytlin,
\textit{``{Kappa-symmetry of superstring sigma model and generalized 10d
  supergravity equations}''},
\textsf{\doiref{10.1007/JHEP06(2016)174}{JHEP~1606,~174~(2016)}},
\texttt{\arxivref{1605.04884}{arxiv:1605.04884}}.

\bibitem{Mikhailov:2012id}
A.~Mikhailov,
\textit{``{Cornering the unphysical vertex}''},
\textsf{\doiref{10.1007/JHEP11(2012)082}{JHEP~1211,~082~(2012)}},
\texttt{\arxivref{1203.0677}{arxiv:1203.0677}}.

\bibitem{Hoare:2015wia}
B.~Hoare and A.~A.~Tseytlin,
\textit{``{Type IIB supergravity solution for the T-dual of the $\eta$-deformed
  AdS$_{5} \times$ S$^{5}$ superstring}''},
\textsf{\doiref{10.1007/JHEP10(2015)060}{JHEP~1510,~060~(2015)}},
\texttt{\arxivref{1508.01150}{arxiv:1508.01150}}.

\bibitem{Kawaguchi:2014qwa}
I.~Kawaguchi, T.~Matsumoto and K.~Yoshida,
\textit{``{Jordanian deformations of the $AdS_5 x S^5$ superstring}''},
\textsf{\doiref{10.1007/JHEP04(2014)153}{JHEP~1404,~153~(2014)}},
\texttt{\arxivref{1401.4855}{arxiv:1401.4855}}.

\bibitem{Matsumoto:2014cja}
T.~Matsumoto and K.~Yoshida,
\textit{``{Integrable deformations of the AdS$_5 \times$S$^5$ superstring and
  the classical Yang-Baxter equation--- \emph{Towards the gravity/CYBE
  correspondence}---}''},
\textsf{\doiref{10.1088/1742-6596/563/1/012020}{J.Phys.Conf.Ser.~563,~012020~(2014)}},
\texttt{\arxivref{1410.0575}{arxiv:1410.0575}}.

\bibitem{vanTongeren:2015soa}
S.~J.~van~Tongeren,
\textit{``{On classical Yang-Baxter based deformations of the AdS$_{5}$ ×
  S$^{5}$ superstring}''},
\textsf{\doiref{10.1007/JHEP06(2015)048}{JHEP~1506,~048~(2015)}},
\texttt{\arxivref{1504.05516}{arxiv:1504.05516}}.

\bibitem{Wulff:2013kga}
L.~Wulff,
\textit{``{The type II superstring to order $\theta^4$}''},
\textsf{\doiref{10.1007/JHEP07(2013)123}{JHEP~1307,~123~(2013)}},
\texttt{\arxivref{1304.6422}{arxiv:1304.6422}}.

\bibitem{Kyono:2016jqy}
H.~Kyono and K.~Yoshida,
\textit{``{Supercoset construction of Yang-Baxter deformed AdS$_5\times$S$^5$
  backgrounds}''},
\texttt{\arxivref{1605.02519}{arxiv:1605.02519}}.

\bibitem{Bedoya:2010qz}
O.~A.~Bedoya, L.~I.~Bevilaqua, A.~Mikhailov and V.~O.~Rivelles,
\textit{``{Notes on beta-deformations of the pure spinor superstring in AdS(5)
  x S(5)}''},
\textsf{\doiref{10.1016/j.nuclphysb.2011.02.012}{Nucl.~Phys.~B848,~155~(2011)}},
\texttt{\arxivref{1005.0049}{arxiv:1005.0049}}.

\bibitem{Hoare:2016ibq}
B.~Hoare and S.~J.~van~Tongeren,
\textit{``{Non-split and split deformations of AdS$_5$}''},
\texttt{\arxivref{1605.03552}{arxiv:1605.03552}}.

\bibitem{Hoare:2016hwh}
B.~Hoare and S.~J.~van~Tongeren,
\textit{``{On jordanian deformations of AdS$_5$ and supergravity}''},
\texttt{\arxivref{1605.03554}{arxiv:1605.03554}}.

\bibitem{Orlando:2016qqu}
D.~Orlando, S.~Reffert, J.-i.~Sakamoto and K.~Yoshida,
\textit{``{Generalized type IIB supergravity equations and non-Abelian
  classical r-matrices}''},
\texttt{\arxivref{1607.00795}{arxiv:1607.00795}}.

\bibitem{Lunin:2005jy}
O.~Lunin and J.~M.~Maldacena,
\textit{``{Deforming field theories with U(1) x U(1) global symmetry and their
  gravity duals}''},
\textsf{\doiref{10.1088/1126-6708/2005/05/033}{JHEP~0505,~033~(2005)}},
\texttt{\arxivref{hep-th/0502086}{hep-th/0502086}}.

\bibitem{Frolov:2005ty}
S.~A.~Frolov, R.~Roiban and A.~A.~Tseytlin,
\textit{``{Gauge-string duality for superconformal deformations of N=4 super
  Yang-Mills theory}''},
\textsf{\doiref{10.1088/1126-6708/2005/07/045}{JHEP~0507,~045~(2005)}},
\texttt{\arxivref{hep-th/0503192}{hep-th/0503192}}.

\bibitem{Frolov:2005dj}
S.~Frolov,
\textit{``{Lax pair for strings in Lunin-Maldacena background}''},
\textsf{\doiref{10.1088/1126-6708/2005/05/069}{JHEP~0505,~069~(2005)}},
\texttt{\arxivref{hep-th/0503201}{hep-th/0503201}}.

\bibitem{Gursoy:2005cn}
U.~Gursoy and C.~Nunez,
\textit{``{Dipole deformations of N=1 SYM and supergravity backgrounds with
  U(1) x U(1) global symmetry}''},
\textsf{\doiref{10.1016/j.nuclphysb.2005.07.023}{Nucl.~Phys.~B725,~45~(2005)}},
\texttt{\arxivref{hep-th/0505100}{hep-th/0505100}}.

\bibitem{Metsaev:2001bj}
R.~R.~Metsaev,
\textit{``{Type IIB Green-Schwarz superstring in plane wave Ramond-Ramond
  background}''},
\textsf{\doiref{10.1016/S0550-3213(02)00003-2}{Nucl.~Phys.~B625,~70~(2002)}},
\texttt{\arxivref{hep-th/0112044}{hep-th/0112044}}.

\bibitem{Metsaev:1998it}
R.~Metsaev and A.~A.~Tseytlin,
\textit{``{Type IIB superstring action in AdS$_5 \times$S$^5$ background}''},
\textsf{\doiref{10.1016/S0550-3213(98)00570-7}{Nucl.Phys.~B533,~109~(1998)}},
\texttt{\arxivref{hep-th/9805028}{hep-th/9805028}}.

\bibitem{Arutyunov:2008if}
G.~Arutyunov and S.~Frolov,
\textit{``{Superstrings on AdS(4) x CP**3 as a Coset Sigma-model}''},
\textsf{\doiref{10.1088/1126-6708/2008/09/129}{JHEP~0809,~129~(2008)}},
\texttt{\arxivref{0806.4940}{arxiv:0806.4940}}.

\bibitem{Stefanski:2008ik}
B.~Stefanski,~jr,
\textit{``{Green-Schwarz action for Type IIA strings on AdS(4) x CP**3}''},
\textsf{\doiref{10.1016/j.nuclphysb.2008.09.015}{Nucl.~Phys.~B808,~80~(2009)}},
\texttt{\arxivref{0806.4948}{arxiv:0806.4948}}.

\bibitem{Gomis:2008jt}
J.~Gomis, D.~Sorokin and L.~Wulff,
\textit{``{The Complete AdS(4) x CP**3 superspace for the type IIA superstring
  and D-branes}''},
\textsf{\doiref{10.1088/1126-6708/2009/03/015}{JHEP~0903,~015~(2009)}},
\texttt{\arxivref{0811.1566}{arxiv:0811.1566}}.

\bibitem{Babichenko:2009dk}
A.~Babichenko, J.~Stefanski,~B. and K.~Zarembo,
\textit{``{Integrability and the AdS$_3$/CFT$_2$ correspondence}''},
\textsf{\doiref{10.1007/JHEP03(2010)058}{JHEP~1003,~058~(2010)}},
\texttt{\arxivref{0912.1723}{arxiv:0912.1723}}.

\bibitem{Sorokin:2011rr}
D.~Sorokin, A.~Tseytlin, L.~Wulff and K.~Zarembo,
\textit{``{Superstrings in AdS(2)xS(2)xT(6)}''},
\textsf{\doiref{10.1088/1751-8113/44/27/275401}{J.~Phys.~A44,~275401~(2011)}},
\texttt{\arxivref{1104.1793}{arxiv:1104.1793}}.

\bibitem{2004math......2433T}
V.~N.~{Tolstoy},
\textit{``{Chains of extended Jordanian twists for Lie superalgebras}''},
\textsf{ArXiv~Mathematics~e-prints~A44,~V.~N.~{Tolstoy}~(2004)},
\texttt{\arxivref{math/0402433}{math/0402433}}.

\bibitem{Stolin1991a}
A.~Stolin,
\textit{``{On rational solutions of Yang-Baxter equation for
  $\mathfrak{sl}(n)$}''},
\textsf{Math.~Scand.~69,~57~(1991)}.

\bibitem{STOLIN1999285}
A.~Stolin,
\textit{``Rational solutions of the classical Yang-Baxter equation and quasi
  Frobenius Lie algebras''},
\textsf{\doiref{http://dx.doi.org/10.1016/S0022-4049(97)00217-X}{Journal~of~Pure~and~Applied~Algebra~137,~285
  ~(1999)}},
\href{http://www.sciencedirect.com/science/article/pii/S002240499700217X}{\texttt{http://www.sciencedirect.com/science/article/pii/S002240499700217X}}.

\bibitem{Gerstenhaber1997}
M.~Gerstenhaber and A.~Giaquinto,
\textit{``Boundary Solutions of the Classical Yang--Baxter Equation''},
\textsf{\doiref{10.1023/A:1007363911649}{Letters~in~Mathematical~Physics~40,~337~(1997)}},
\href{http://dx.doi.org/10.1023/A:1007363911649}{\texttt{http://dx.doi.org/10.1023/A:1007363911649}}.

\bibitem{Stolin1991b}
A.~Stolin,
\textit{``{Constant solutions of Yang-Baxter equation for $\mathfrak{sl}(2)$
  and $\mathfrak{sl}(3)$}''},
\textsf{Math.~Scand.~69,~81~(1991)}.

\bibitem{Lichnerowicz1988}
A.~Lichnerowicz and A.~Medina,
\textit{``On Lie groups with left-invariant symplectic or K{\"a}hlerian
  structures''},
\textsf{\doiref{10.1007/BF00398959}{Letters~in~Mathematical~Physics~16,~225~(1988)}},
\href{http://dx.doi.org/10.1007/BF00398959}{\texttt{http://dx.doi.org/10.1007/BF00398959}}.

\bibitem{Ovando2006}
G.~Ovando,
\textit{``Four dimensional symplectic Lie algebras.''},
\textsf{Beiträge~zur~Algebra~und~Geometrie~47,~419~(2006)},
\href{http://eudml.org/doc/226718}{\texttt{http://eudml.org/doc/226718}}.

\bibitem{Patera:1973yn}
J.~Patera, P.~Winternitz and H.~Zassenhaus,
\textit{``{The Maximal Solvable Subgroups Of The $Su(p,q)$ Groups And All
  Subgroups Of $Su(2,1)$}''},
\textsf{\doiref{10.1063/1.1666820}{J.~Math.~Phys.~15,~1378~(1974)}}.

\bibitem{Patera1973b}
J.~Patera, P.~Winternitz and H.~Zassenhaus,
\textit{``The maximal solvable subgroups of SO(p,q) groups''},
\textsf{\doiref{http://dx.doi.org/10.1063/1.1666559}{Journal~of~Mathematical~Physics~15,~1932~(1974)}},
\href{http://scitation.aip.org/content/aip/journal/jmp/15/11/10.1063/1.1666559}{\texttt{http://scitation.aip.org/content/aip/journal/jmp/15/11/10.1063/1.1666559}}.

\bibitem{Alday:2005ww}
L.~F.~Alday, G.~Arutyunov and S.~Frolov,
\textit{``{Green-Schwarz strings in TsT-transformed backgrounds}''},
\textsf{\doiref{10.1088/1126-6708/2006/06/018}{JHEP~0606,~018~(2006)}},
\texttt{\arxivref{hep-th/0512253}{hep-th/0512253}}.

\bibitem{Elitzur:1994ri}
S.~Elitzur, A.~Giveon, E.~Rabinovici, A.~Schwimmer and G.~Veneziano,
\textit{``{Remarks on nonAbelian duality}''},
\textsf{\doiref{10.1016/0550-3213(94)00426-F}{Nucl.~Phys.~B435,~147~(1995)}},
\texttt{\arxivref{hep-th/9409011}{hep-th/9409011}}.

\bibitem{Bergshoeff:1995as}
E.~Bergshoeff, C.~M.~Hull and T.~Ortin,
\textit{``{Duality in the type II superstring effective action}''},
\textsf{\doiref{10.1016/0550-3213(95)00367-2}{Nucl.~Phys.~B451,~547~(1995)}},
\texttt{\arxivref{hep-th/9504081}{hep-th/9504081}}.

\bibitem{Green:1996bh}
M.~B.~Green, C.~M.~Hull and P.~K.~Townsend,
\textit{``{D-brane Wess-Zumino actions, t duality and the cosmological
  constant}''},
\textsf{\doiref{10.1016/0370-2693(96)00643-0}{Phys.~Lett.~B382,~65~(1996)}},
\texttt{\arxivref{hep-th/9604119}{hep-th/9604119}}.

\bibitem{Hassan:1999bv}
S.~F.~Hassan,
\textit{``{T duality, space-time spinors and RR fields in curved
  backgrounds}''},
\textsf{\doiref{10.1016/S0550-3213(99)00684-7}{Nucl.~Phys.~B568,~145~(2000)}},
\texttt{\arxivref{hep-th/9907152}{hep-th/9907152}}.

\bibitem{Hashimoto:1999ut}
A.~Hashimoto and N.~Itzhaki,
\textit{``{Noncommutative Yang-Mills and the AdS / CFT correspondence}''},
\textsf{\doiref{10.1016/S0370-2693(99)01037-0}{Phys.~Lett.~B465,~142~(1999)}},
\texttt{\arxivref{hep-th/9907166}{hep-th/9907166}}.

\bibitem{Maldacena:1999mh}
J.~M.~Maldacena and J.~G.~Russo,
\textit{``{Large N limit of noncommutative gauge theories}''},
\textsf{\doiref{10.1088/1126-6708/1999/09/025}{JHEP~9909,~025~(1999)}},
\texttt{\arxivref{hep-th/9908134}{hep-th/9908134}}.

\bibitem{Matsumoto:2014gwa}
T.~Matsumoto and K.~Yoshida,
\textit{``{Integrability of classical strings dual for noncommutative gauge
  theories}''},
\textsf{\doiref{10.1007/JHEP06(2014)163}{JHEP~1406,~163~(2014)}},
\texttt{\arxivref{1404.3657}{arxiv:1404.3657}}.

\bibitem{vanTongeren:2015uha}
S.~J.~van~Tongeren,
\textit{``{Yang–Baxter deformations, AdS/CFT, and twist-noncommutative gauge
  theory}''},
\textsf{\doiref{10.1016/j.nuclphysb.2016.01.012}{Nucl.~Phys.~B904,~148~(2016)}},
\texttt{\arxivref{1506.01023}{arxiv:1506.01023}}.

\end{thebibliography}

\end{document}